\PassOptionsToPackage{unicode}{hyperref}
\PassOptionsToPackage{hyphens}{url}
\PassOptionsToPackage{dvipsnames,svgnames,x11names}{xcolor}
\documentclass[12pt, ]{article}
\usepackage{xcolor}
\usepackage[margin=1in]{geometry}
\usepackage{amsmath,amssymb}
\usepackage{iftex}
\ifPDFTeX
  \usepackage[T1]{fontenc}
  \usepackage[utf8]{inputenc}
  \usepackage{textcomp} %
\else %
  \usepackage{unicode-math} %
  \defaultfontfeatures{Scale=MatchLowercase}
  \defaultfontfeatures[\rmfamily]{Ligatures=TeX,Scale=1}
\fi
\usepackage{lmodern}
\ifPDFTeX\else
\fi
\IfFileExists{upquote.sty}{\usepackage{upquote}}{}
\IfFileExists{microtype.sty}{%
  \usepackage[]{microtype}
  \UseMicrotypeSet[protrusion]{basicmath} %
}{}
\makeatletter
\@ifundefined{KOMAClassName}{%
  \IfFileExists{parskip.sty}{%
    \usepackage{parskip}
  }{%
    \setlength{\parindent}{0pt}
    \setlength{\parskip}{6pt plus 2pt minus 1pt}}
}{%
  \KOMAoptions{parskip=half}}
\makeatother
\makeatletter
\ifx\paragraph\undefined\else
  \let\oldparagraph\paragraph
  \renewcommand{\paragraph}{
    \@ifstar
      \xxxParagraphStar
      \xxxParagraphNoStar
  }
  \newcommand{\xxxParagraphStar}[1]{\oldparagraph*{#1}\mbox{}}
  \newcommand{\xxxParagraphNoStar}[1]{\oldparagraph{#1}\mbox{}}
\fi
\ifx\subparagraph\undefined\else
  \let\oldsubparagraph\subparagraph
  \renewcommand{\subparagraph}{
    \@ifstar
      \xxxSubParagraphStar
      \xxxSubParagraphNoStar
  }
  \newcommand{\xxxSubParagraphStar}[1]{\oldsubparagraph*{#1}\mbox{}}
  \newcommand{\xxxSubParagraphNoStar}[1]{\oldsubparagraph{#1}\mbox{}}
\fi
\makeatother

\usepackage{longtable,booktabs,array}
\usepackage{calc} %
\usepackage{etoolbox}
\makeatletter
\patchcmd\longtable{\par}{\if@noskipsec\mbox{}\fi\par}{}{}
\makeatother
\IfFileExists{footnotehyper.sty}{\usepackage{footnotehyper}}{\usepackage{footnote}}
\makesavenoteenv{longtable}
\usepackage{graphicx}
\makeatletter
\newsavebox\pandoc@box
\newcommand*\pandocbounded[1]{%
  \sbox\pandoc@box{#1}%
  \Gscale@div\@tempa{\textheight}{\dimexpr\ht\pandoc@box+\dp\pandoc@box\relax}%
  \Gscale@div\@tempb{\linewidth}{\wd\pandoc@box}%
  \ifdim\@tempb\p@<\@tempa\p@\let\@tempa\@tempb\fi%
  \ifdim\@tempa\p@<\p@\scalebox{\@tempa}{\usebox\pandoc@box}%
  \else\usebox{\pandoc@box}%
  \fi%
}
\def\fps@figure{htbp}
\makeatother

\usepackage[round]{natbib}
\usepackage{setspace}
\usepackage[english]{babel}

\usepackage{amsfonts}   %

\usepackage{tabularx}
\usepackage{dcolumn}
\usepackage{adjustbox}
\usepackage{multirow}
\usepackage{multicol}
\usepackage{rotating}
\usepackage[para,online,flushleft]{threeparttable}
\usepackage{rccol}
\usepackage{fltpoint}
\usepackage{amssymb}
\usepackage{tabularray}
\UseTblrLibrary{booktabs}
\SetTblrStyle{caption-tag}{font=\bfseries}
\SetTblrStyle{caption-sep}{font=\bfseries}
\usepackage{siunitx}

\usepackage{colortbl}

\usepackage{lscape}
\usepackage{pdflscape}
\usepackage{placeins}    %
\usepackage{pdfpages}

\usepackage[capposition=top]{floatrow}
\usepackage{adjustbox}
\usepackage{capt-of}
\usepackage{tikz}
\usetikzlibrary{decorations.pathreplacing, math}
\usepackage{pgffor}
\usepackage{pgfplots}
\pgfplotsset{compat=1.18}
\usepgfplotslibrary{fillbetween}

\usepackage{ragged2e}
\usepackage{changepage}
\usepackage{csquotes}
\usepackage{soul}          %
\usepackage{framed}
\usepackage{verbatim}
\usepackage{comment}
\usepackage{marvosym}      %

\usepackage{enumitem}
\setlist{topsep=0pt, leftmargin=*, itemsep=-1pt, parsep=0pt}

\usepackage{chngcntr}
\usepackage{tocbibind}     %
\usepackage{hyperref}
\usepackage{cleveref}      %

\newcolumntype{P}[1]{>{\centering\arraybackslash}p{#1}}

\def\sym#1{\ifmmode^{#1}\else\(^{#1}\)\fi}

\newenvironment{itemize*}
  {\begin{itemize}
    \setlength{\itemsep}{0pt}%
    \setlength{\topsep}{0pt}
    \setlength{\partopsep}{0pt}
    \setlength{\parsep}{0pt}
    \setlength{\parskip}{0pt}}
  {\end{itemize}}

\newenvironment{enumerate*}
  {\begin{enumerate}
    \setlength{\itemsep}{0pt}
    \setlength{\parsep}{0pt}}
  {\end{enumerate}}
\usepackage{booktabs}
\usepackage{caption}
\usepackage{longtable}
\usepackage{colortbl}
\usepackage{array}
\usepackage{anyfontsize}
\usepackage{multirow}
\makeatletter
\@ifpackageloaded{caption}{}{\usepackage{caption}}
\AtBeginDocument{%
\ifdefined\contentsname
  \renewcommand*\contentsname{Table of contents}
\else
  \newcommand\contentsname{Table of contents}
\fi
\ifdefined\listfigurename
  \renewcommand*\listfigurename{List of Figures}
\else
  \newcommand\listfigurename{List of Figures}
\fi
\ifdefined\listtablename
  \renewcommand*\listtablename{List of Tables}
\else
  \newcommand\listtablename{List of Tables}
\fi
\ifdefined\figurename
  \renewcommand*\figurename{Figure}
\else
  \newcommand\figurename{Figure}
\fi
\ifdefined\tablename
  \renewcommand*\tablename{Table}
\else
  \newcommand\tablename{Table}
\fi
}
\@ifpackageloaded{float}{}{\usepackage{float}}
\floatstyle{ruled}
\@ifundefined{c@chapter}{\newfloat{codelisting}{h}{lop}}{\newfloat{codelisting}{h}{lop}[chapter]}
\floatname{codelisting}{Listing}

\makeatother
\makeatletter
\@ifpackageloaded{tikz}{}{\usepackage{tikz}}
\makeatother
\makeatletter
\@ifpackageloaded{caption}{}{\usepackage{caption}}
\@ifpackageloaded{subcaption}{}{\usepackage{subcaption}}
\makeatother
\usepackage{bookmark}
\IfFileExists{xurl.sty}{\usepackage{xurl}}{} %
\hypersetup{
  pdftitle={Labor Market Returns to Education and Intra-family Educational Inequality: Theory and Experimental Evidence},
  pdfauthor={Esther Gehrke; Michelle González Amador},
  colorlinks=true,
  linkcolor={blue},
  filecolor={Maroon},
  citecolor={Blue},
  urlcolor={Blue},
  pdfcreator={LaTeX via pandoc}}

\begin{document}

\begin{titlepage}
\begin{singlespace}

\title{%
 Labor Market  Returns to Education and Intra-family Educational Inequality%
  \thanks{%
    We thank seminar participants at Wageningen University for valuable feedback. We thank Dr. Salvador Ruiz-Correa, Ixchelt Cerda, Paty Ruiz, Jorge Cuendia for excellent fieldwork assistance. This lab-in-the-field obtained ethical approval by the Ethics
    Committee of the University of Wageningen (IRB approval date: 08.08.2025) and was pre-registered with the AEA
    Trial Registry (Trial ID: AEARCTR-0017183). Funding by the Dutch Research
    Organization (Grant number: VI.Veni.221E.045) is gratefully acknowledged.%
  }\\[0.4em]
  {\large\normalfont
    Theory and Experimental Evidence}%
}

\author{%
  Esther Gehrke%
    \thanks{%
      Wageningen University \& Research, The Netherlands.
      E-mail: \href{mailto:esther.gehrke@wur.nl}{esther.gehrke@wur.nl}.%
    }%
  \and
  Michelle González Amador%
    \thanks{%
      Wageningen University \& Research, The Netherlands.
      E-mail: \href{mailto:michelle.gonzalezamador@wur.nl}{michelle.gonzalezamador@wur.nl}.%
    }%
}

\date{2 November 2026}

\maketitle             %

\begin{abstract}
\noindent
How do labor market conditions shape educational inequality within- and between families?
A growing body of evidence documents convex wage returns to education in labor markets across the world. We develop a theoretical model that demonstrates that earnings convexity can increase or decrease between-sibling differences in investments and outcomes depending on the parameter of parental inequality aversion. 
We recover the distribution of this structural parameter from a highly-incentivized lab-in-the-field experiment with 246 parents in San Luis Potosí, Mexico, in which token allocations determine children's access to an AI course. Parents make allocation decisions across six scenarios that vary convexity in the payoff function and endowment-investment substitutability. We identify three  types of parents: Rawlsian parents display perfect inequality aversion (12.6\%), interior parents that are moderately inequality tolerant (76.8\%), and Utilitarian parents who maximize total payoffs (10.6\%). While these three groups respond to earnings convexity in heterogeneous ways, on average, greater convexity increases the share allocated to the high-endowed child, thereby widening within-family inequality.

\end{abstract}



\thispagestyle{empty}  %
\end{singlespace}
\end{titlepage}

\section{Introduction}\label{sec-intro}

Parental investments in children's education have consequential implications for adult life outcomes in terms of health, income, and well-being. In low income settings, where resource constraints are  binding, parents must carefully choose how to allocate resources between their children: Preferences for equal  outcomes (e.g. future earnings or well-being) between children have to be weighted against the cost of compensating differences in initial endowments. How these choices are made determines between-sibling and between-family educational inequality, and bears consequences for social mobility. 

Seminal work by \cite{behrman1982parental} has argued that whether parents decide to compensate  or reinforce initial differences in endowments is fully governed by the parental inequality aversion parameter. More recent work has sought to distinguish preferences for equality in inputs from preferences over outcomes \citep{berry2025not}, and to determine how these preferences interact with beliefs about child ability and the complementarity between ability and inputs in the human capital production function \citep{giannola2024parental}. However, much less is known about the role of labor market conditions, despite evidence suggesting that wage returns to education are convex in a variety of settings \citep{autor2014skills, acemoglu2011chapter, gethin2025}, which may disproportionally raise the cost of equalizing child outcomes.

In this paper, we study how convexity in returns to education interacts with parental preferences in determining human capital investments, outcomes, and between-sibling inequalities. To this end,  we  first develop a theoretical model that extends prior work by specifically allowing for non-linearity in the earnings function. 
We derive analytical and numerical solutions to demonstrate that the parental decision to compensate or reinforce pre-existing ability differentials depends not only on parental inequality aversion, but crucially also on the ease by which such compensation can take place. In our framework, this ease is governed by the degree of convexity in the earnings function and by the complementarity between ability and investment in the production of child human capital.

We then implement a highly incentivized lab-in-the-field experiment with 246 parents of at least two high school-enrolled children in San Luis Potosí, Mexico, to identify the structural parameters that govern the parental investment decision. 
Across six scenarios, parents allocate nine tokens between two of their children. Because the children’s identities remain concealed throughout the task, parents must base their allocations solely on the performance information provided.
One scenario is selected at random for implementation, so parents' choices determine actual educational investments rather than hypothetical outcomes. The tokens translate into access to up to 15 sessions of an Artificial Intelligence (AI) course for their children, designed and implemented in collaboration with local educational partners. The experiment varies two features that determine how tokens translate into AI sessions across scenarios: The substitutability between each child's initial endowment (assigned by the experimenters) and parental investments (tokens), and the convexity of the earnings function. This design creates non-linear within-parent variation in the cost of equalizing outcomes between siblings. When initial endowment and tokens are more substitutable, compensating the low-endowment child becomes less costly; when the earnings function is more convex, the cost of compensation (in terms of foregone sessions to the high-endowment child) increases disproportionally. Observing how the same parent allocates tokens across these six scenarios allows us to recover the parent's inequality aversion parameter in a constant elasticity of substitution (CES) value function, which we designate by $\rho$.

We recover the sample distribution of the inequality-aversion parameter using the two-step generalized method of moments (GMM) \citep{hansen1982large,newey1994large}. The model fits the allocation data well: The mean absolute error is $0.84$ tokens on a 0--9-token allocation scale, so the typical prediction lies within one token of the observed choice, and the correlation between predicted and observed allocations is $0.88$.

From the estimated parameter, we identify three classes of parents. The first group, comprising 12.6\% of parents, displays Rawlsian preferences ($\hat{\rho} \to -\infty$). Their choices are consistent with perfect inequality aversion: Parents are willing to sacrifice substantial payoffs in order to equalize outcomes across children (in some scenarios, this implies foregoing almost 90\% of potential payoffs). %
At the other extreme, 10.6\% of parents are classified as Utilitarians, as they display no inequality aversion at all ($\hat{\rho}=1$). These parents choose payoff-maximizing allocations throughout and allocate sharply toward the high-performing child. %
The remaining 76.8\% of parents have interior levels of inequality aversion. In this interior group, the median inequality aversion parameter $\hat{\rho}$ is $0.50$, with an interquartile range (IQR) of $[-1.12,0.79]$. %

The reduced-form estimates validate the behavioral patterns implied by the structural parameters: Higher substitutability between endowment and investment shifts allocations towards more compensation. On average, higher convexity in payoffs shifts parental allocations toward the high-endowment child, and as predicted this effect is entirely driven by non-Rawlsian parents. %

Our findings have direct implications for the design and evaluation of policies aimed at reducing educational inequality. 
As returns to education become increasingly convex, the price that inequality averse parents pay in terms of foregone earnings increases over-proportionally, leaving their children with substantially lower education and earnings, individually and in total, than children of less inequality averse parents. These lower outcomes would affect educational investments across generations, amplifying educational inequalities between families (but not within), and inhibiting social mobility.  By contrast, parents who are (moderately) inequality tolerant, respond to earnings convexities by allocating even more resources towards the higher-endowed child, increasing between-sibling inequalities in education and earnings, with moderate effects on aggregate inequality. Furthermore, if families are able to redistribute these resources through transfers or support towards the next generation, then within-family inequality may carry lower welfare losses than between-family inequality. 
Determining how large these different groups are in the overall population of parents, and potentially even identifying them, has implications for the targeting of educational resources and the design of policies.

This paper contributes to the literature in two ways. \textit{First}, we extend existing theoretical models with which to study intra-household resource allocation problems. \cite{behrman1982parental} conceptualized parental resource allocation as the outcome of a trade-off between maximizing the sum of child earnings and equity concerns.\footnote{Crucially, the earnings function maps educational outcomes to labor-market earnings.} Subsequent work has expanded this model to highlight, \textit{inter alia}, the importance of dynamic complementarities in the human capital production function \citep{heckman2014mosso, giannola2024parental}. To the best of our knowledge, this is the first paper to derive how the properties of the earnings function shape parental resource allocation between children. 

\textit{Second}, we contribute to the empirical evidence that seeks to identify   whether parents reinforce or compensate pre-existing endowment differentials between their children. So far, this literature is inconclusive: \citet{aizer2012production} exploit exogenous variation from the initial launch of Head Start in 1966 and show that parents, through their investments, tend to reinforce initial endowment differences between siblings. Similar evidence of sibling rivalry in low-income contexts \citep{carneiro2023families, barrera2023compensation, berman2025sweet, giannola2024parental} supports the notion that parents indeed reinforce existing differences, rather than compensate them. 
By contrast, other studies have found that parents compensate differences in health or educational outcomes between their children \citep{yi2015early}. 

To the best of our knowledge, we are among the first to link parental allocations not only to preferences over between-sibling inequality but also to systematic variation in the cost of compensation driven by the convexity in returns to education, as well as by the degree of substitutability between endowments and investments. Most closely related to our paper is \cite{giannola2024parental}, who estimates the parameter of parental inequality aversion in India using a hypothetical survey experiment, and finds that parents are relatively inequality tolerant. In contrast to that paper, we recover inequality aversion from a highly-incentivized allocation task, in which parents need to make choices for their own children, rather than for two hypothetical children. It is perhaps unsurprising that parents are more inequality averse when the allocation concerns their own children as opposed to another family's children.    

The remainder of this paper proceeds as follows.  \Cref{sec-theory} introduces the theoretical framework and derives model predictions. \Cref{sec-experiment} describes the experimental design. \Cref{sec-data} presents the setting and data. \Cref{sec-results} details the structural estimation strategy, presents the main results, and reports reduced-form evidence. \Cref{sec-conclusion} concludes.

\section{Theoretical Framework}\label{sec-theory}
\subsection{Setup}\label{subsec-setup}

As in standard models of intra-household allocation
\citep[e.g.,][]{behrman1982parental, heckman2014mosso},
parents derive utility from their children's adult-life outcomes, and optimize educational resource allocation subject to a budget constraint. We consider a household with two children indexed $i \in \{1, 2\}$, each characterized by innate ability $a_i$ and receiving parental investment $x_i$.\footnote{Replacing innate ability by early investments generalizes this model to one that incorporates dynamic complementarities.} In line with the literature, we normalize total educational expenditures to one,  such that $x_2=1-x_1$, and focus only on how those expenditures are split between the two children.

Child $i$'s human capital is produced with a constant elasticity of substitution (CES) technology:
\begin{equation}
H_i(a_i,x_i) = (\gamma a_i^{\sigma} + (1-\gamma) x_i^{\sigma})^{1/\sigma}.
\end{equation}
The parameter $\sigma \in [- \infty, 1]$ governs the complementarity between ability and investments. The weight $\gamma \in (0, 1)$ captures the importance of ability relative to investments in determining the human capital outcome. 

The earnings accruing to child $i$ are a function of their human
capital:
\begin{equation}
E_i(a_i,x_i) = \alpha_{\theta} H_i^\theta,
\end{equation}
where $\alpha_{\theta} > 0$ scales the average return to human capital and $\theta$ governs the convexity in the earnings function. When $\theta = 1$, the earnings function is linear in human capital; when $\theta > 1$, it is  convex. Convexity in earnings introduces an efficiency motive for concentrating resources on the higher-ability child (i.e.~reinforcing existing ability differences) that is absent under linearity.

Parents choose investment levels $\{x_1, x_2\}$ to maximize a CES value function defined over their children's adult-life outcomes:
\begin{equation} 
V = [\mu ( E_1(a_1, x_1)^\phi)^\rho + (1-\mu) (E_2(a_2,x_2)^\phi)^\rho]^{\frac{1}{\rho}}.
\label{eq:valuefunc}
\end{equation}
Three preference parameters guide parental allocations. The weight $\mu \in (0,1)$ captures child-specific weights, such as gender or birth-order bias, but also idiosyncratic preferences for one child over the other. The parameter $\phi > 0$ captures the curvature in utility over each individual child's earnings (parental risk aversion): A parent with $\phi < 1$ places diminishing marginal utility on child earnings as earnings increase, and is therefore less willing to concentrate resources in a single child.\footnote{With $\phi=1$ the problem simplifies to one in which the parental value function can be written over children's earnings, as in \cite{behrman1982parental}. We introduce $\phi$ in order to illustrate that inequality aversion, as well as risk aversion  over earnings could drive equal allocations.} The parameter $\rho$ is the between-child inequality aversion parameter. As $\rho \to -\infty$, the value function converges to the Rawlsian case and parents always seek to equalize outcomes. As $\rho \to 1$, parents display no inequality aversion, and purely maximize the (weighted) sum of child outcomes. Between these extremes, parents trade off efficiency and equity concerns, proportionally to $\rho$.

\subsection{Analytical Solutions}\label{subsec-analyticalsolutions}

We derive the equilibrium log investment  ratio from the first-order condition of the parental optimization problem. In the most general case, this gives:
\begin{equation}
\ln \left(\frac{x_1}{x_2}\right) = \frac{1}{1-\sigma} \ln \left(\frac{\mu}{1-\mu}\right) + \frac{\theta \phi \rho - \sigma}{1-\sigma} \ln \left(\frac{H_1}{H_2} \right).
\label{eq:general-allocation}
\end{equation}
While this expression illustrates that the allocative decisions are determined by $\mu$, $\theta$, $\phi$, $\rho$, and $\sigma$, the direction of the response is not immediately obvious because $H_i$ is itself an outcome of the investment $x_i$. 

To obtain a closed-form analytical solution, we restrict our attention to the case of a Cobb-Douglas production technology ($\sigma \to 0$). We further assume \textit{equal concern} (i.e. $\mu=0.5$) for analytical tractability. The log investment ratio becomes:\footnote{Under the additional restrictions of no risk aversion over child earnings ($\phi = 1$), and linear returns ($\theta = 1$), the solution reduces to the canonical framework of \citet{behrman1982parental}: Parents with $\rho > 0$ reinforce existing ability gaps, concentrating resources on the higher-ability child; while parents with $\rho < 0$ compensate them. The formal derivation of \cref{eq:cd-allocation} is presented in \Cref{app:analytical-solution}.}
\begin{equation}
\ln \left(\frac{x_1}{x_2}\right) =  \frac{\theta  \gamma  \phi  \rho }{1-  \left(1-\gamma\right) \theta \phi  \rho} \ln \! \left(\frac{a_1}{a_2}\right).
\label{eq:cd-allocation}
\end{equation}
This expression reveals how parental resource allocation responds to ability differentials as a function of preferences ($\rho, \phi$) and the cost of compensation ($\gamma, \theta$). 

\begin{figure}
\centering{
\input{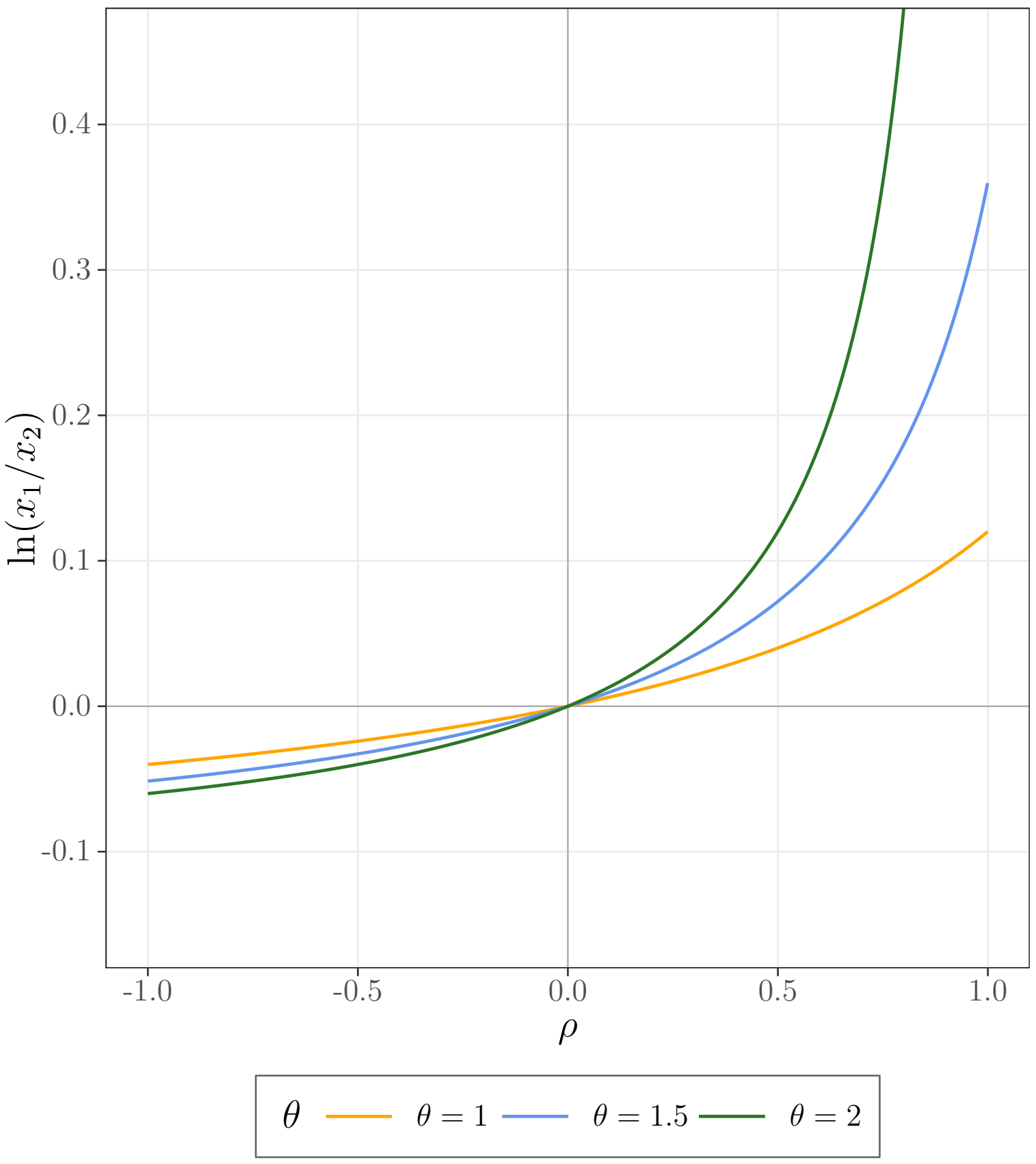}}
\caption{\label{fig-theta-cd}Effect of Convexity on Allocation Decisions --- Analytical Solutions
}
\floatfoot{\textit{Notes}: This figure plots the log investment ratio $\ln(x_1/x_2)$ as function of inequality aversion ($\rho$) at varying levels of convexity in earnings ($\theta$).}
\end{figure}

\Cref{fig-theta-cd} illustrates the role of variation in $\rho$ and $\theta$ in allocative decisions for the case of $a_1=0.53$, $a_2=0.47$, $\gamma=1/2$,  $\mu=1/2$ and $\phi=1$. Because $a_1>a_2$, $\ln(a_1/a_2)>0$.  For any  positive log investment ratio ($\ln(x_1/x_2)>0$),  the parent reinforces existing inequalities in ability by investing more in the more endowed child; for any $\ln(x_1/x_2)<0$, the parent compensates initial inequalities by investing more in the less endowed child. Under a Cobb-Douglas human capital production technology, the parent switches from compensating to reinforcing behavior at $\rho=0$ for any level of $\theta$  \citep[in line with][]{behrman1982parental}. Interestingly, $\theta$ amplifies both types of behavior: Compensating parents ($\rho<0$) compensate even more to counteract the inequality-increasing effect of convexity in the earnings function. By contrast, reinforcing parents ($\rho>0$) respond to increases in $\theta$ by amplifying their reinforcing behavior, further amplifying between-sibling inequality. 

\subsection{Numerical Solutions for General Values of $\sigma$}\label{subsec-numericalsolutions}

Beyond the Cobb-Douglas case ($\sigma \to 0$), the first-order condition of \cref{eq:valuefunc} has no closed-form solution, so we solve for the decision rule numerically.\footnote{In the Cobb-Douglas case, $\ln(H_1/H_2)$ is linear in $\ln(a_1/a_2)$ and in $\ln(x_1/x_2)$, which yields the closed-form allocation rule in \cref{eq:cd-allocation}. For $\sigma \neq 0$, the investment ratio enters inside the CES aggregator, implying that the sign of the log human capital ratio $\ln(H_1/H_2)$ in $\ln(a_1/a_2)$ is ambiguous.
} 
The prediction of interest is how the allocation of investments responds to changes in the convexity of the earnings function across values of $\rho$ and $\sigma$. 

\begin{figure}
    \centering
    \includegraphics[width=0.75\linewidth]{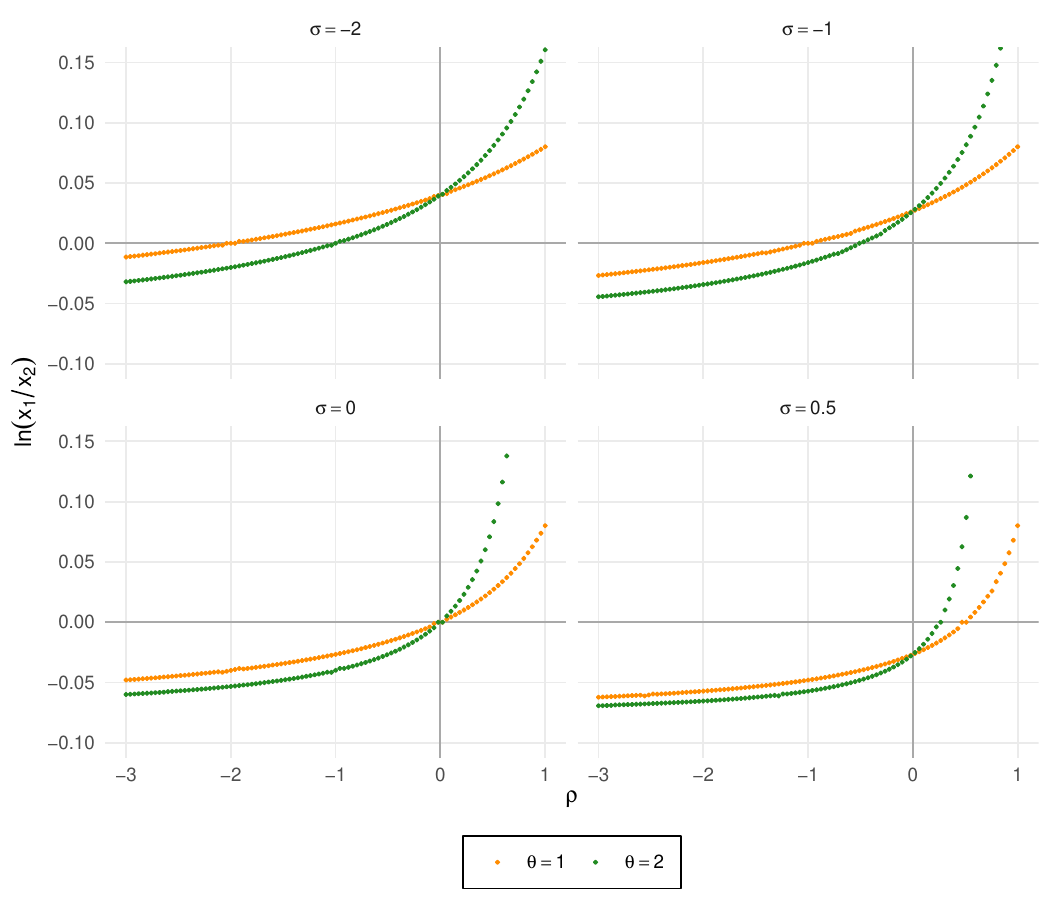}
    \caption{Effect of Convexity on Allocation Decisions --- Numerical Solutions}
    \label{fig:predictions_lndiff}
    \floatfoot{\textit{Notes}: The figure shows how $\ln{(x_{\text{1}}/x_{\text{2}}})$ varies with $\rho$ across values of $\sigma$, comparing $\theta = 1$ and $\theta = 2$.}     
\end{figure}

\Cref{fig:predictions_lndiff} replicates \Cref{fig-theta-cd} for four representative values of $\sigma$, i.e. {-2, -1, 0, 0.5}. For ease of comparison, we keep $a_1=0.53$, $a_2=0.47$, $\gamma=1/2$,  $\mu=1/2$ and $\phi=1$.\footnote{We show simulations in which we relax the assumption of $\mu=1/2$ and $\phi=1$ in \Cref{fig:relax-phi-mu}. The key insight is that $\phi<1$ affects the intensity but not the direction of any effect of $\theta$, as does $\mu>1/2$; only weighting the lower-ability child ($\mu<1/2$) can reverse the direction of the $\theta$ effect.}
Three important insights emerge: \textit{First}, for linear earnings ($\theta =1$), the switching point from compensating to reinforcing behavior is where $\rho=\sigma$. At levels of inequality aversion for which $\rho<\sigma$, parents compensate; at $\rho>\sigma$ parents reinforce. This shift in strategy reflects the varying cost of compensating for ability differentials: The higher the complementarity between ability and investments in producing child human capital ($\sigma \to - \infty$), the more costly compensation becomes in terms of foregone human capital of the more endowed child.  This cost is high enough that even moderately inequality-averse parents $(\rho \in (-2, 0))$ find it optimal to reinforce under linear earnings. 

\textit{Second}, convexity in earnings shifts this threshold. For negative values of $\sigma$, i.e. ability and investments are complements, the threshold $\rho$ at which parents switch from compensatory to reinforcing behavior becomes less negative (approaches zero) with increasing convexity. For positive values of $\sigma$, i.e. ability and inputs are substitutes, the threshold $\rho$ at which parents switch from compensatory to reinforcing behavior becomes smaller with increasing convexity. 

\textit{Third,} along the range of $\rho$, the slope of the log investment ratio is steeper in $\rho$ for higher levels of $\theta$.  This implies that compensating behavior is amplified for sufficiently inequality averse parents, who react to increases in earnings inequality by compensating even more. At the same time, reinforcing behavior is also amplified for sufficiently inequality tolerant parents, who embrace the incentives given by the convexity in the earnings function to further concentrate their resources in the high-endowment child. Parents at intermediate levels of inequality aversion may switch strategies as $\theta$ increases, or display declining reinforcement or compensation.\footnote{Similar insights can be generated for the log ratio of human capital outcomes $\ln(H_1/H_2)$, as shown in \Cref{fig:predictions_lnHdiff}.}    

The overall effect of an increase in the convexity in earnings on between-sibling inequality in educational investments is therefore ambiguous, and depends on parental inequality aversion as well as on parental perceptions about the complementarity between ability and investments in the production of child human capital. The finding that distinct values of $\rho$ generate different responses to increasing convexity across different $\sigma$ values, is the cornerstone of our experimental design.

\section{Experimental Design and Procedures}\label{sec-experiment}

We design and implement a lab-in-the-field experiment with two siblings and their parents to identify three parental preference parameters: $\rho$, $\phi$, and $\mu$. %
We are primarily interested in the identification of the $\rho$ parameter, for which we developed a highly-incentivized experimental task. We recover $\phi$ from survey instruments with parents, and $\mu$ from validated survey instruments implemented with children.

\subsection{Overview of the Experimental Task}\label{subsec-design}

For the experimental task, both children first completed a short questionnaire, which included a 12-item Raven's Progressive Matrices test, independently of each other. More details on the questionnaire's content are given in Section \ref{subsec-procedure}. Thereafter, one of their parents was interviewed.\footnote{Parents were recruited via an invitation letter delivered to them by the participating schools; the letter invited parents and their two children to participate in the (international) study, highlighting the academic nature of the study, the benefits for the families and for society and, importantly, a surprise gift for their kids. The surprise, revealed only during the experimental task, were AI courses.} 

For the experimental task, parents were told that their children had completed a problem-solving test and that one sibling outperformed the other. They were then informed that their children had the opportunity to participate in a purpose-built AI course designed for high school students (grades 10--12), co-developed by the research team and local experts from the Instituto Potosino de Investigación Científica y Tecnológica (IPICyT). The course consisted of 15 standalone sessions delivered outside school hours, combining foundational AI concepts with real-world examples and applications. The sessions were specifically described as standalone units, such that any additional session has the same value in terms of additional knowledge imparted. In our description, we also stressed the interactive nature of the course, such that parents understood that their children could not teach each other.\footnote{Parents were informed at the time of the experiment that the course would begin some months later, and that the opportunity was not expected to recur, a framing designed to prevent expectations about the younger child being able to participate in a similar course in the future from distorting allocation decisions.} We then informed parents that due to budget constraints, not every child could participate in the full course, but only for a limited number of sessions. This number was to be determined by their children's performance in the above-mentioned test, in combination with a lottery element.\footnote{The lottery element referred to the random selection of the allocation scenario whose associated parental choices would determine course participation.} In addition, they were given the opportunity to influence the number of sessions each child could enroll in by allocating nine tokens between their children. Parents were asked to make these allocations across six distinct scenarios, one of which would be implemented at random after the interview. 

To prevent choices from being influenced by knowing which child would benefit from a given allocation, child identities were not disclosed: Throughout the task, children were referred to only as \textcolor{orange}{child orange} and \textcolor{Green}{child green}, with \textcolor{orange}{child orange} being the relatively better performing child. This blinding also ensures that $\mu \approx 0.5$ by design, isolating the estimated $\rho$ from child-specific preference weights. 
Furthermore, parents were never shown their children's actual Raven's scores to minimize the possibility of parental backlash. Instead, parents could only see how their allocations translated into final payoffs.\footnote{After the experiment, parents were revealed which allocation was drawn and implemented, but would have to ask their children to find out how many sessions each of them were allowed to enroll in.}

\begin{figure}
\centering{
\pandocbounded{\includegraphics[keepaspectratio]{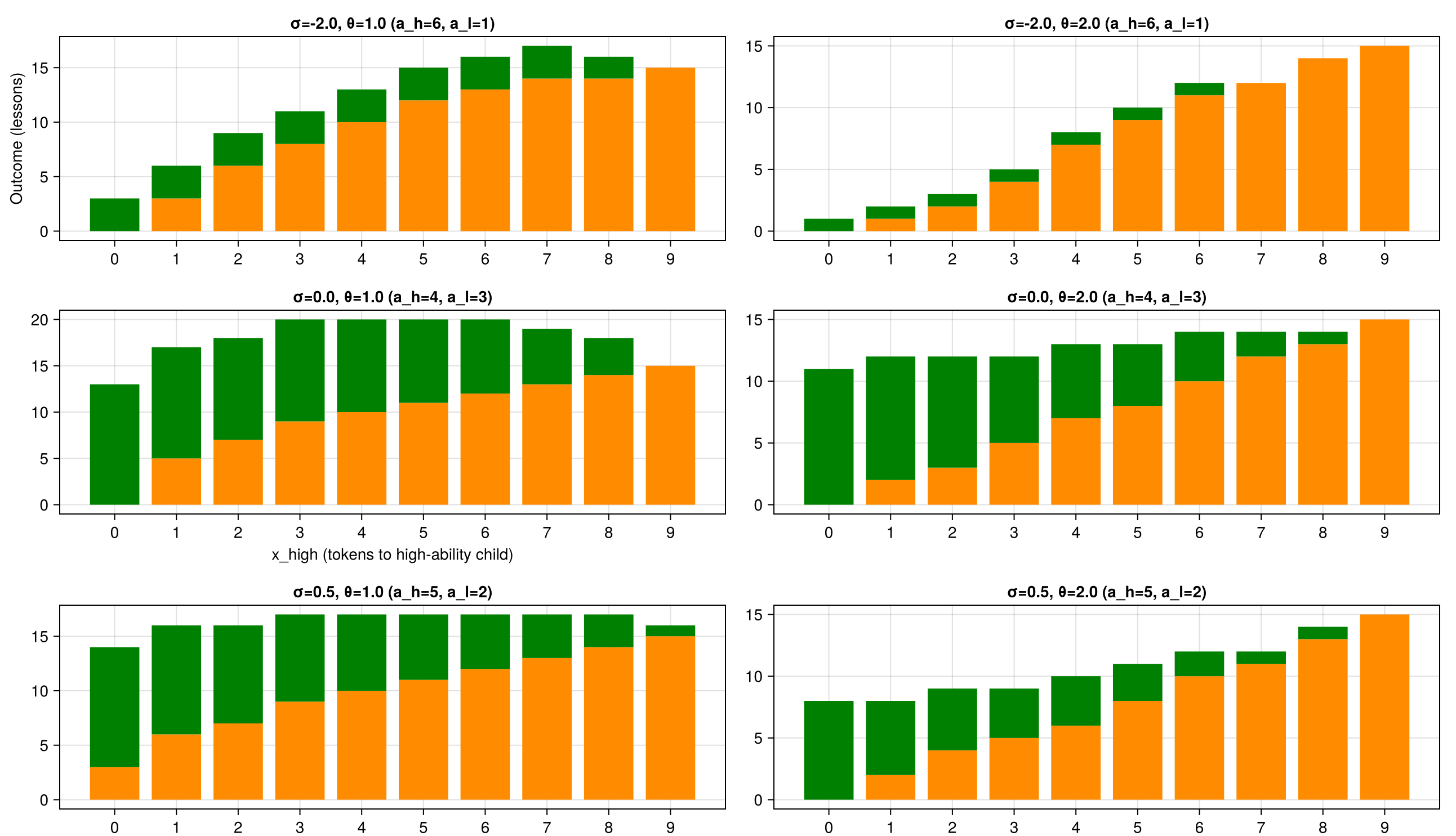}}
}
\caption{\label{fig-scenarios}The Six Experimental Scenarios}
\floatfoot{\textit{Notes}: Each bar shows the payoff outcome (AI course
sessions) for \textcolor{orange}{child orange} and \textcolor{Green}{child green}, stacked.}
\end{figure}

Each scenario was presented graphically on a tablet. Scenarios were shown in random order, and for each scenario the left-right positions of \textcolor{orange}{child orange} and \textcolor{Green}{child green} are independently randomized. Parents could allocate the nine tokens between their two children by moving a slider; bars updated in real time to display the payoff outcome for each child under that scenario's payoff function and the parent's choice (cf.~\Cref{fig-scenarios}). To explain the existence of different scenarios, parents were told that this was a way to keep students from realizing how well they did in comparison to others, as this would generate random variation between students.

\subsection{Scenario Selection}\label{subsec-identification}

The scenarios are chosen to maximize precision in a within-subject design. We introduce exogenous variation in $\theta$ and $\sigma$ across scenarios to separate the production environment from parental preferences, and trace how allocations respond to changes in earnings convexity across different levels of $\sigma$. We implement a $2 \times 3$ factorial design, crossing two levels of  payoff convexity ($\theta \in \{1, 2\}$) with three levels of complementarity between endowment (ability in the theoretical framework) and investment ($\sigma \in \{-2, 0, 0.5\}$), as shown in \Cref{tab:factorial-design}. 

\begin{table}[h!t]
    \centering
        \caption{The $2 \times 3$ Repeated-Measures Factorial Design}
    \small
\begin{threeparttable}
    \begin{tabular}{l |c|c|c}
    \toprule
     & \multicolumn{3}{c}{\textbf{Substitutability between endowment and inputs}} \\
   \textbf{Payoffs} & \textbf{Low ($\sigma=-2)$} & \textbf{Cobb-Douglas} & \textbf{High  ($\sigma=0.5$)} \\
    \midrule
         \textbf{Linear} & large gap & small gap & mid gap \\
         \textbf{Convex} & large gap & small gap & mid gap \\
    \bottomrule     
    \end{tabular}
\end{threeparttable}

    \vspace{0.4em}
    \begin{minipage}{0.85\linewidth}
        \scriptsize \textit{Notes:} The gap sizes correspond to a 6--1 token difference for the large gap, 5--2 for the mid-sized gap, and 4--3 token difference for the small gap.
    \end{minipage}
    \label{tab:factorial-design}
\end{table}

Across these six scenarios, payoffs were \textit{not} based on the children's actual Raven's scores, which partly reflect prior investment history and would confound the estimate of $\rho$, but instead are researcher-controlled endowment values whose size is calibrated to the level of $\sigma$.\footnote{Not disclosing true performance is a common strategy in this type of experiment \citep[see e.g.,][]{berry2025not}. Also note that the Raven's ranking is used solely by the researchers to assign course sessions at payoff implementation.}  Specifically, a small endowment gap (4-3) in the Cobb-Douglas scenario, medium (5-2) for $\sigma = 0.5$, and large (6-1) for $\sigma = -2$. These values were chosen to minimize the estimated error around $\rho$ in a series of Monte Carlo Simulations.

\subsection{Implementation and Procedures}\label{subsec-procedure}

Sessions were conducted in a dedicated room at each participating school and lasted approximately 90 minutes per household. Each session involved one parent and their two co-enrolled children, attended by at least two trained enumerators; the number of enumerators per session varied with school size, reaching up to 16 when multiple families attended simultaneously. To prevent any prior communication between parents and children about the task, children were interviewed first; parents were invited to join only after all children at that school had completed their survey.\footnote{In cases where school scheduling created overlap between parent and student arrival times, parents were seated in a separate room and had no contact with the children until after the children's stage was complete.}

The children completed a short survey independently and simultaneously. The child survey covered questions related to aspirations and expectations, a validated depression module \citep[CES-10, Spanish]{RuizGrosso2012}, as well as a validated parental bias module, the Sibling Inventory of Differential Experience \citep[SIDE,][]{daniels1985differential}, which we use to  construct $\mu$. The SIDE module asks children to compare how each parent treats them relative to their sibling across items on affection, control, and favoritism; we scale the resulting measure so that $\mu=0.5$ denotes equal perceived treatment. The survey was followed by a 12-item Raven's Progressive Matrices test to measure fluid intelligence and pattern recognition. Raven's tests are commonly used as a language-independent proxy for cognitive ability. Before taking the test, children were informed that the test was incentivized and that their payoff would be revealed at the end of the data collection period, but were not told that their parent would play any role in what they received.\footnote{Parents were interviewed on the same day as their children, so children were informed that the payoff was an AI course within the course of the day, but they were also informed that the number of sessions they could enroll in was still to be decided.}

Subsequently, an enumerator administered the parent survey, which covered demographic characteristics, educational aspirations and expectations for each child, and subjective beliefs about each child's academic performance relative to their peers, beliefs about labor market returns to education, and a Holt--Laury risk aversion elicitation with respect to future labor market earnings of their children.%

Following the survey, the enumerator introduced the allocation task. Two practice rounds --- using a grocery shopping exercise instead of the allocation of AI course sessions between their children, and serving simultaneously as comprehension checks (see \Cref{subsec-comprehension}) --- familiarized parents with the slider interface and the real-time graphical feedback before the incentivized task began. The six experimental scenarios were then presented in a randomized order; for each scenario, the enumerator read a standardized description and then passed the tablet to the parent, who could move the slider to their preferred allocation in private. 

After all six allocations were recorded, parents completed an exit survey, which included self-reported equity preferences and some questions concerning parental perceptions of the AI course. Crucially, the exit survey included a willingness to pay (WTP) task used to construct an auxiliary $\phi$ measure. Although the description of the course emphasized that each of the 15 sessions was self-contained and could be understood without necessarily attending any more sessions --- thereby minimizing curvature in the demand for the course --- this does not \textit{per se} ensure that $\phi=1$ for the AI sessions. In order to obtain individual level $\phi$ values, parental willingness to pay (WTP) for an additional course session was elicited at three hypothetical baseline levels of attained sessions.\footnote{Note that the curvature in the demand ($\phi$) need not be the same for child payoffs in the experimental task as for adult-life earnings of their children. We therefore included two separate tasks to recover two distinct structural parameters. } %
 The parents were then informed that they would receive an email with the course start dates and exact payoffs per child at the end of the data collection period.

Data collection ran from November 2025 to July 2026 (with interruptions during the school break from December to March). To incentivize participation without revealing the nature of the payoff, i.e.~disclosing the AI course in advance could have introduced experimenter demand effects, both parents and children were informed by the school that participation in the research project would automatically enter them into a raffle.\footnote{Children were eligible to win a Samsung smartphone (of three total); parents were entered into a separate raffle for MXN \$25,000 (ca. USD \$1,450). 
The raffle was held (online) in March 2026 for parents interviewed until then; raffle winners were notified by email and through their children's schools. For parents interviewed after that date, participation was incentivized with a fixed MXN \$500 each.}

\subsection{Comprehension Check and Sample Restrictions}\label{subsec-comprehension}

To verify that parents understood the allocation task before engaging in the incentivized scenarios, we administered two practice rounds using a neutral store framing: Parents were asked to allocate nine tokens between two familiar supermarkets --- Soriana ({\color{orange}{orange}}) and Walmart ({\color{Green}{green}}) --- which sell identical goods at identical prices, geographically located side by side. By design, this framing holds $\mu = 0.5$ and $\phi = 1$; there is no reason to prefer one store over the other except that one values the tokens more, making the payoff-maximizing allocation (all nine tokens to the store with the higher token value) transparent and independent of any preference for the store itself. Both practice rounds used the same payoff parameters ($\sigma = 1$), ($\theta = 2$). A parent who placed all tokens on the higher-value store demonstrated that they understood the mapping from allocations to payoffs; we use this as a binary comprehension indicator. Full details of the practice scenario script are provided in \Cref{app:comprehension}.

Of the $N = 278$ parents who participated in the experiment, $251$ passed at least one round. $27$ parents failed both checks and are excluded from the primary estimation sample, on the grounds that their choices cannot be reliably interpreted as preference-revealing under the structural model. We further exclude $5$ parents that exhibit no variation in their token allocations, which is indicative of disengagement or inattention.\footnote{These five parents chose nine tokens for {\color{orange}{child orange}} throughout. Given that this was not the payoff-maximizing choice in three of the six scenarios, we have to conclude that they were not paying attention to the task.} The final estimation sample consists of $246$ parents and their two children.

\section{Setting and Data}\label{sec-data}

\subsection{Sample and Setting}\label{subsec-setting}

The experiment was conducted in San Luis Potosí (SLP), a state in central Mexico. SLP is a useful setting for studying parental investments because returns to education are both large and visible. Around one in five working-age residents holds a tertiary degree \citep{inegi2020censo}, yet the wage premium to tertiary education remains substantial. In SLP, tertiary graduates earn 47.9\% higher wages than upper-secondary graduates.\footnote{We use ENOE 2025 microdata \citep{inegi2025enoe} to predict log wages at different levels of education, controlling for age, age squared, and gender.}

\begin{figure}
    \centering
    \includegraphics[width=0.7\linewidth]{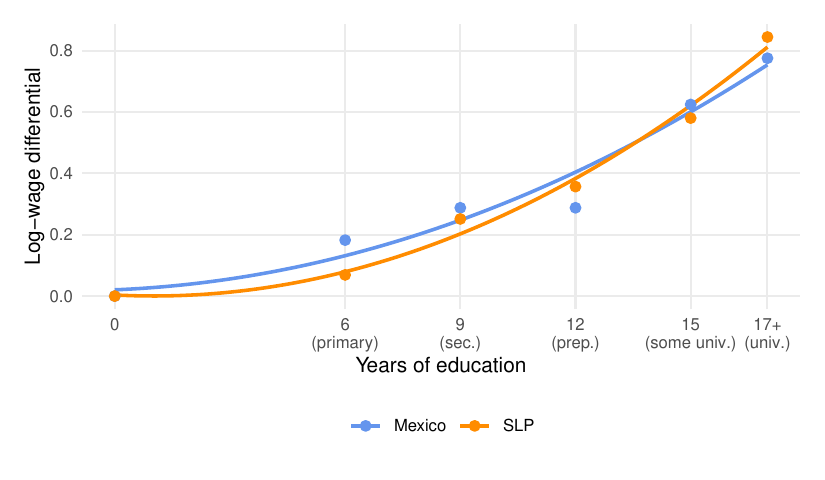}
    \caption{Convexity in Earnings: Mexico and SLP}
    \label{fig-convexity}
    {\scriptsize\emph{Notes:} Wages are monthly labor earnings for q1--q4 in 2025. Points are survey-weighted medians using ENOE's quarterly expansion weight, \texttt{fac\_tri}.\par}
\end{figure}

These returns are not only high but convex. \Cref{fig-convexity} plots the raw log-wage differential by years of education for Mexico and SLP. In both cases, the wage gain from moving into completed tertiary education is disproportionally larger than the wage gain at lower attainment levels. The returns in SLP  closely track the national pattern and are, if anything, even more convex.

We recruited urban families from the SLP metropolitan area and nearby urban municipalities.\footnote{The sampling area includes peri-urban municipalities with large high schools, as well as Ciudad Valles, the largest urban center in the neighboring Huasteca region.} Eligible families had at least two children, ages ca. 15--18, simultaneously enrolled in the same high school (\emph{preparatoria}). The same-school restriction was imposed to facilitate data collection as well as the implementation of the AI course. An important feature of the design is that we only include parents who are able to afford to send two children to high school, which may not be true for the average parent in the country or even in the state.

Parents were recruited through 16 high schools. Schools served as contact points for reaching eligible families; the design does not exploit school-level variation. Participating schools distributed invitations to potentially eligible families and interested families were assigned a session date and time. Because schools do not maintain systematic records of co-enrolled siblings, the eligible population is unknown and formal take-up rates cannot be computed. The sample includes $N=278$ participating families, of whom $N=246$ form the final estimation sample.\footnote{A pre-registered power analysis supporting the target sample size is provided in \Cref{app:power}.}

\subsection{Descriptive Statistics}\label{subsec-data}

The final dataset combines child responses with parent responses. Descriptive statistics for the estimation sample are reported in \Cref{tab:summary-estimation-sample}. 

The average parent in our sample is 43 years old, and 68\% of the participating parents are the children's mothers. More than half of the parents (67.5\% of mothers and 66.3\% of fathers) have completed at least upper-secondary education. Households have 5 members on average, and mean reported monthly household income is approximately MXN $20,905$. Educational aspirations are high: Parents aspire to university or more for 95.5\% of children and expect university or more for 91.5\%. 

In terms of child characteristics and child outcomes, it is worth noting that children are on average 16 years old, with a mean age gap of 1.5 years. The students' average Raven score is 8.3 on a range from 0 to 12 (one point per correct item), and the average  Raven gap between siblings is 1.9 points. Children also perceive their parents to have little systematic preference for one child over the other: The average SIDE score, with regards to both mothers and fathers, is around 0.5, indicating equal treatment.

\section{Empirical Strategy and Results}\label{sec-results}

\subsection{Estimating the Structural Model Parameter}\label{subsec-estimation}

We identify the inequality aversion parameter $\rho$ from the structural model and estimate it separately for each parent $i \in \{1, \ldots, 246\}$ via two-step generalized method of moments (GMM) estimator \citep{hansen1982large, newey1994large, chamberlain1982multivariate}. For each parent, we construct six moment conditions $\mathbf{m}_i(\rho) \in \mathbb{R}^6$ as the differences between observed allocations $x^{\text{obs}}_{ij}$ and model-predicted allocations $x^*_j(\rho,\phi_i)$ across the experimental scenarios $j = 1, \ldots, 6$. For a given $\rho$, we solve the household optimization problem numerically, compute $\mathbf{m}_i(\rho)$, and search for the value minimizing the weighted distance to observed moments. Note that because $\rho_i$ and $\phi_i$ enter the allocation rule through their product, $\rho_i\phi_i$, the experimental choices do not separately identify the two parameters. We therefore estimate $\rho_i$ conditional on an independently elicited, parent-specific $\phi_i$.\footnote{Our main specification uses the individual level $\phi$ elicited in the WTP exercise, winsorized at the 5th and 95th percentiles. Results are robust to setting $\phi = 1$ for all parents, and to using the raw $\phi$ values.}

Stage one estimates $\hat{\rho}_{1,i}$ by minimizing $L_1(\rho) = \mathbf{m}_i(\rho)' \mathbf{I} \, \mathbf{m}_i(\rho)$, where $\mathbf{I} \in \mathbb{R}^{6 \times 6}$ is the identity matrix, yielding a consistent estimator. Stage two implements the standard GMM weighting matrix \citep{hansen1982large}, setting $W = S^{-1}$ where $S = \tfrac{1}{N}\sum_{i=1}^{N} \hat{\mathbf{r}}_i \hat{\mathbf{r}}_i'$ is the empirical moment-covariance matrix constructed from stage-one residuals pooled across parents. Pooling across parents is necessary because six scenarios per parent are insufficient to estimate a parent-specific $6 \times 6$ covariance matrix \citep{hansen1996finite}; assuming a common moment-covariance structure across parents is justified by the experimental design, since every parent faces the same six $(\theta, \sigma)$ scenarios. Re-estimation minimizes $L_2(\rho) = \mathbf{m}_i(\rho)' W \, \mathbf{m}_i(\rho)$, where $W$ is the inverse cross-parent second-moment matrix of the stage-one residuals. The weighting accounts for differences in residual variance and covariance across the six scenario moments.\footnote{Stage two gives less direct weight to the linear-return scenarios and more direct weight to the convex-return scenarios, especially ($\sigma=0.5, \theta=2$). The estimated weights are reported in \Cref{app:model-fit}, \Cref{tab:stage2-weights}.}  The parameter $\rho_i$ is identified by the extent to which variation in each parent's observed allocation choices across the six scenarios aligns with the model-predicted choices as $\rho$ varies.

\subsection{Estimated Distribution of Inequality
Aversion}\label{subsec-rho}

The GMM estimator yields individual-level estimates of $\rho$ for all 246 parents in the estimation sample. We first classify estimates into three preference types: Rawlsian, interior, and Utilitarian. This classification is useful because the two extremes have different interpretations from interior estimates. At the lower boundary, preferences approach the Rawlsian case ($\hat{\rho} \to  - \infty$): Choices are consistent with parents maximizing the outcome of the worst-off child. We classify a parent as Rawlsian if their token allocation choices exactly match the integer maximin vector (\texttt{1,4,3,1,4,3}) in the slider scenarios, or if their estimated $\hat{\rho}$ lies at the lower search boundary.\footnote{The main search grid is constrained to $\hat{\rho}\in[-80,1]$ for computational efficiency. In lower-bound sensitivity checks, we expand the grid to $\hat{\rho}\in[-300,1]$ and $\hat{\rho}\in[-500,1]$. Estimates for the Rawlsian group follow the grid expansions into the negative space, leading us to conclude that in this group $\hat{\rho} \to -\infty$.} This group consists of 31 parents (12.6\% of the sample, see \Cref{tab:pref-categories}). 
At the other extreme, $\hat{\rho}=1$, parents %
consistently maximize the sum of child payoffs (Utilitarians). These parents are willing to trade-off the two children's payoffs one-for-one. However, when several allocations yield the same maximum total payoff, parents in this group often select the most equal allocation in the set, suggesting that equality is still an important consideration for them. %
This group consists of 26 parents (10.6\% of the sample).

\begin{table}[htbp]
\centering

\begin{tabular}{llrr}
\toprule
Category & Definition & $N$ & Share\\
\midrule
Rawlsian & $\hat{\rho}\rightarrow -\infty$ & 31 & 12.6\%\\
Interior & $-\infty < \hat{\rho}<1$ & 189 & 76.8\%\\
Utilitarian & $\hat{\rho} = 1$ & 26 & 10.6\%\\
\bottomrule
\end{tabular}
\caption{Distribution of Inequality-Aversion Types}
\label{tab:pref-categories}
\end{table}

The last group of parents displays interior levels of $\hat{\rho}$, and consists of 189 parents (76.8\% of the sample). Among the interior group, the distribution is centered around moderate inequality tolerance. The interior  $\hat{\rho}$ has a median of $0.50$, a mean of $-1.64$, and the interquartile range is $[-1.12,0.79]$ (see \Cref{tab:rho-sumstats}). \Cref{fig-logces} plots the distribution of the log elasticity of substitution, $\ln(\hat{\eta})$, with $\eta=1/(1-\rho)$, for the estimation sample, documenting the relative importance of  the three types of parents. As can be seen, the distribution is indeed trimodal, reflecting the three types of parents described in \Cref{tab:pref-categories}. The mode of the interior group ($\ln(\hat{\eta})= 1.3$) corresponds to a level of $\hat{\rho}$ of 0.72 which is close to the median level of $\hat{\rho}$ in the interior group.\footnote{The results are  robust to alternative specifications of $\phi$ and $\mu$: Point estimates vary only marginally, while the three $\hat{\rho}$ classifications and the $(\sigma,\theta)$ patterns remain unchanged, see \Cref{tab:rho-sensitivity-phi-mu}.}

\noindent
\begin{minipage}[t]{0.36\linewidth}
\vspace{0pt}
\centering

\captionof{table}{Summary statistics for interior $\rho$}
\label{tab:rho-sumstats}

\renewcommand{\arraystretch}{0.8}
\resizebox{0.75\linewidth}{!}{%
  
\begin{tabular}{lr}
\toprule
Statistic & Value\\
\midrule
$N$ & 189\\
\addlinespace[0.2em]
Mean & $-1.64$\\
\addlinespace[0.2em]
SD & $6.12$\\
\addlinespace[0.2em]
Median & $0.50$\\
\addlinespace[0.2em]
IQR & $[-1.12,0.79]$\\
\addlinespace[0.2em]
Range & $[-57.136,0.986]$\\
\bottomrule
\end{tabular}
}
\end{minipage}\hfill%
\begin{minipage}[t]{0.62\linewidth}
\vspace{0pt}
\centering

\captionof{figure}{Log elasticity of substitution}
\label{fig-logces}

\includegraphics[width=\linewidth]{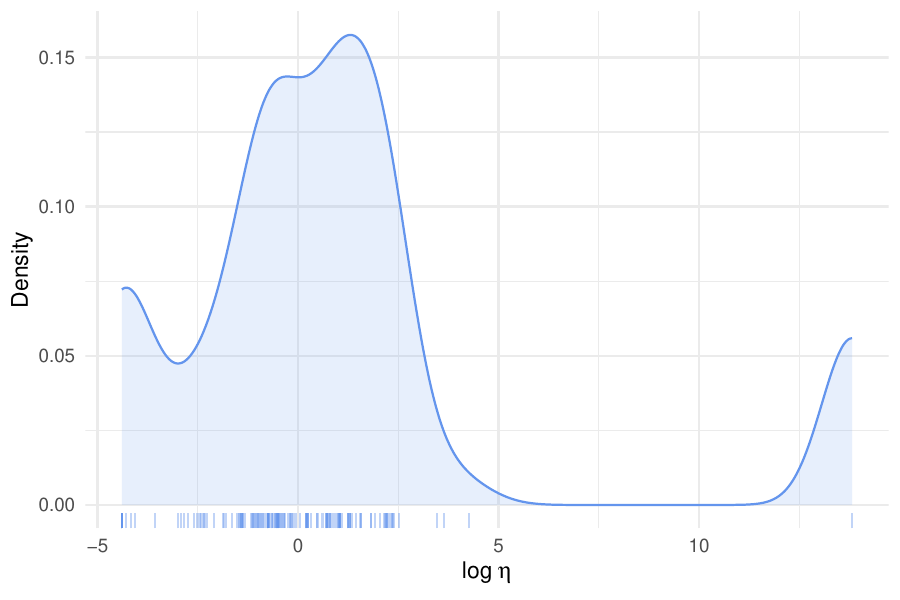} %
\end{minipage}%

\subsection{Model Fit}\label{subsec-fit}

We assess goodness of fit across all $246 \times 6 = 1{,}476$ allocations, by scenario, and by parent. Pooled mean absolute (MAE) and root mean squared (RMSE) errors are $0.84$ and $1.28$ tokens, respectively, on a 0--9 token scale. The mean residual is $-0.18$, indicating slight over-prediction of allocations to the high-endowment child. Predicted and observed allocations are strongly correlated ($r=0.88$; $R^2=0.76$), and the median parent-level RMSE is $1.17$ tokens. Scenario-level fit is strongest in the convex cases ($\theta=2$; rows 4--6 of \Cref{tbl-scenario-fit}), where prediction errors are lowest and the model slightly under-predicts allocations to the high-endowment child. Within these cases, fit is best at very high and very low levels of substitutability.\footnote{This pattern is consistent with sharper incentives narrowing the range of attractive allocations and thereby making parents’ choices more predictable. Additional goodness-of-fit diagnostics are reported in \Cref{app:model-fit}.} Overall, the model explains substantial variation with errors that are modest relative to the nine-token budget. %

\begin{table}[htbp]
\centering
\caption{Scenario-specific fit}
\label{tbl-scenario-fit}
\begin{threeparttable}

\begin{tabular}{ccrrr}
\toprule
$\theta$ & $\sigma$ & MAE & RMSE & Mean residual\\
\midrule
1 & $-2$ & 0.85 & 1.25 & $-0.67$\\
1 & $0$ & 0.91 & 1.24 & $-0.43$\\
1 & $0.5$ & 1.51 & 1.93 & $-0.70$\\
2 & $-2$ & 0.53 & 0.96 & $+0.21$\\
2 & $0$ & 0.72 & 1.06 & $+0.30$\\
2 & $0.5$ & 0.53 & 0.94 & $+0.20$\\
\bottomrule
\end{tabular}
\begin{tablenotes}[flushleft]
\small
\item \textit{Notes:} Stage-2 weighted GMM, $n = 246$ parents per scenario. Mean residual is the average of $x^{\text{obs}} - x^*$; negative values indicate the model over-predicts the allocation to the high-endowment child.
\end{tablenotes}
\end{threeparttable}
\end{table}

\subsection{Reduced-Form Evidence}\label{sec-reduced-form}

As a complement to the structural estimates, we also document the reduced-form relationship between the experimental parameters ($\sigma$, $\theta$) and observed allocations. These regressions provide evidence that parents respond to the incentive structure in theoretically predicted directions. All specifications are estimated by OLS (linear-in-means and linear probability models) with order of display fixed effects. Standard errors are clustered at the parent level throughout. Identification rests on within-parent cross-scenario variation in the experimental parameters.

We estimate three sets of specifications, corresponding to hypotheses H1--H3 from the pre-analysis plan. Results are reported in \Cref{tab:reduced-form}. Recall that the estimation sample is restricted to parents who passed the slider comprehension check, and varied their token allocations across the six scenarios ($N=246$).

\begin{table}[!htbp]
\centering
\small
\sisetup{group-separator={,}, group-minimum-digits=4}
\begin{talltblr}[
  caption = {Reduced-Form Evidence: Effects of Substitutability ($\sigma$) and Return Convexity ($\theta$) on Token Allocations},
  label   = {tab:reduced-form},
  note{}  = {\textit{Notes:} Dependent variable reported in column header. Standard errors clustered at the parent level are shown in parentheses. One-sided test p-values are reported in \Cref{tab:hypothesis-tests-final}. \\ * $p<0.1$, ** $p<0.05$, *** $p<0.01$.}
]{
  colspec = {Q[l,3.2cm] Q[c,1.85cm] Q[c,1.85cm] Q[c,1.85cm] Q[c,1.85cm] Q[c,1.85cm]},
  rowsep = 1pt,
  colsep = 2pt,
  hline{7} = {1-6}{solid, 0.05em, black},
}
\toprule
Outcome: & $x_{\text{high}}$ & reinforce & $x_{\text{high}}$ & reinforce & payoffmax \\
 & (1) & (2) & (3) & (4) & (5) \\
\midrule
$\sigma$ & \num{-0.236}*** & \num{-0.115}*** &  &  &  \\
 & (\num{0.049}) & (\num{0.008}) &  &  &  \\
$\theta$ &  &  & \num{1.882}*** & \num{0.260}*** & \num{-0.297}*** \\
 &  &  & (\num{0.128}) & (\num{0.023}) & (\num{0.024}) \\
Obs. & \num{1476} & \num{1476} & \num{1476} & \num{1476} & \num{1476} \\
Households & \num{246} & \num{246} & \num{246} & \num{246} & \num{246} \\
Dep. var. mean & \num{5.257} & \num{0.483} & \num{5.257} & \num{0.483} & \num{0.611} \\
Within $R^2$ & \num{0.024} & \num{0.104} & \num{0.252} & \num{0.114} & \num{0.173} \\
Display-order FE & \checkmark & \checkmark & \checkmark & \checkmark & \checkmark \\
\bottomrule
\end{talltblr}
\end{table}

\medskip

\textbf{H1 - Effect of substitutability on allocation behavior:} As $\sigma$ increases, investment substitutes more easily for endowment, reducing the opportunity cost (in terms of foregone payoffs for the high endowment child) of compensating behavior. The threshold $\bar{\rho}$ above which parents switch from compensating to reinforcing therefore rises with $\sigma$ (cf.~\Cref{fig:predictions_lndiff}).
H1a tests whether higher substitutability shifts the continuous allocation $x_{\text{high}}$ away from the higher-endowed child; H1b tests whether it reduces the probability of reinforcement.

The estimates in \Cref{tab:reduced-form} support this prediction. A one-unit increase in $\sigma$ reduces the allocation to the higher-endowed child by 0.24 tokens (column 1). The estimate is statistically significant under both a two-sided test and the preregistered directional test ($p<0.001$ in both cases; see \Cref{tab:hypothesis-tests-final}). The binary reinforcement probability test provides complementary evidence: A one-unit increase in $\sigma$ decreases the probability of reinforcing behavior by 11.5 percentage points (column 2, $p<0.001$). Thus, greater substitutability between endowments and investments shifts choices in the predicted compensatory direction and reduces the probability that parents reinforce the initial advantage of the higher-endowed child.
As a robustness check, we estimate the same relationships after excluding Rawlsian parents in \Cref{tab:reduced-form-robust}, panels A and B. For both outcomes, the observed effect increases in absolute magnitude, consistent with the notion that Rawlsian parents would not respond to changes in $\sigma$ given that they fully compensate endowment differences throughout.  

\textbf{H2 - Effect of convexity on allocation behavior:}
As $\theta$ increases, returns to investment become more convex, raising the payoff that can be achieved from concentrating resources in the higher-endowed child. H2 tests whether this change in the incentive structure shifts parental allocations. Because the theoretical direction depends on the parent's underlying inequality aversion, we treat this hypothesis as non-directional. H2a tests the continuous allocation $x_{\text{high}}$, while H2b tests the probability of reinforcement.

The estimates in \Cref{tab:reduced-form} show that, on average, convexity substantially increases reinforcement. Moving from linear to convex returns raises the allocation to the higher-endowed child by 1.88 tokens (column 3, $p<0.001$). It also increases the probability that parents reinforce endowment differences by 26 percentage points (column 4, $p<0.001$). These results provide robust evidence that, on average, parents in this sample respond to convex returns by shifting resources toward the child with the higher baseline endowment. Given that the direction of the effect of $\theta$ should vary with parental inequality aversion, we also examine whether the convexity response differs by parents’ estimated $\hat{\rho}$ type. \Cref{tab:h2-rho-type} shows that the positive average effect is concentrated among interior and Utilitarian parents. Convex returns increase $x_{\text{high}}$ by 2.02 tokens among interior parents and by 3.16 tokens among Utilitarian parents. By contrast, the estimate for Rawlsian parents is $-0.02$ tokens and statistically indistinguishable from zero. The corresponding increases in the probability of reinforcement are 28.8 percentage points for interior parents and 37.2 percentage points for Utilitarian parents, while the estimated response among Rawlsian parents is essentially zero. This pattern is consistent with Rawlsian parents maintaining their strongly compensatory allocation strategy when returns become more convex. 

\textbf{H3 - Payoff maximization under convex returns:}
H3 tests whether increasing convexity reduces the probability that parents choose the aggregate-payoff-maximizing allocation. 
In our experimental scenarios, a higher $\theta$ (keeping $\sigma$ constant) shifts the total-payoff maximum toward the extreme allocation in which all tokens are assigned to the higher-endowed child (see \Cref{fig-scenarios}). 
However, only Utilitarian parents would be willing to tolerate such a high level of inequality in child outcomes. For the remaining parents, this allocation would be too unequal  to choose it, even if it maximizes aggregate payoffs. We estimate the average response in the full sample and report the model-derived restriction excluding parents that were classified as Utilitarians ($\hat{\rho}=1$) in \Cref{tab:reduced-form-robust}.

The estimate in \Cref{tab:reduced-form} strongly supports H3. In the full sample of 246 parents, moving from linear to convex returns lowers the probability of choosing the aggregate-payoff-maximizing allocation by 29.7 percentage points (column 5, $p<0.001$). The estimate increases in magnitude to 35.1 percentage points when the 26 Utilitarian parents are excluded ($N=220$; see \Cref{tab:reduced-form-robust}, panel C). This decline indicates that convexity dramatically raises the cost of equalizing outcomes: When the payoff-maximizing choice requires concentrating investment on the higher-endowed child, many parents instead choose allocations that sacrifice aggregate earnings to achieve a more equal \emph{outcome} distribution across siblings.\footnote{The main results are robust to including all parents in the analysis, including those who failed the comprehension checks and those who always allocated all tokens to the higher-endowed child, see \Cref{tab:reduced-form-all-parents}.}

Taken together, the reduced-form results show that parents respond systematically to the experimentally varied incentive structure. On average, higher substitutability shifts token allocations in a compensatory direction and reduces the probability of reinforcement. Higher convexity shifts resources toward the higher-endowed child, with this response concentrating among interior and Utilitarian parents and essentially no response detectable among Rawlsian parents. At the same time, convexity reduces the probability of parents choosing the aggregate-payoff-maximizing allocation, with an even larger decline when Utilitarian parents are excluded.

\section{Conclusion}\label{sec-conclusion}
In this paper, we study how convex returns to education shape parental investment decisions across siblings. Building on the canonical allocation framework of \citet{behrman1982parental}, we develop a model in which parents trade off efficiency against equity, and show that allocative decisions depend not only on parental inequality aversion, but also on the shape of the earnings function and the substitutability between ability and investment in the production of human capital. 

Using an incentivized lab-in-the-field experiment with parents and two co-enrolled children in San Luis Potosí, Mexico, we estimate parent-level inequality aversion, $\rho$, and categorize parental behavior into three empirically distinct types. A non-trivial share of parents, 12.6\%, behave as if they are strongly unwilling to substitute one child's payoff for the other's (Rawlsian parents); a smaller share, 10.6\%, behaves close to payoff maximization (Utilitarian parents); and the remaining 76.8\% have interior estimates, centered on moderate inequality tolerance. The reduced-form evidence confirms the model's core prediction; that is, when investment and endowment are more substitutable, parents shift in a compensatory direction; when returns become more convex, parents shift resources toward the higher-endowed child, but not uniformly. While the majority of parents respond  by investing even more in the higher-endowed child, thereby amplifying within-family educational inequality, a non-negligible group of parents sacrifice substantial payoffs in order to preserve equal outcomes between their children. 

These findings contribute to the literature on intra-household allocation in two ways. \textit{First}, to our knowledge, this is the first paper to place the earnings function at the center of the parental allocation problem, complementing work that emphasizes inequality aversion over inputs rather than outcomes \citep{berry2025not} and the interaction between beliefs, ability, and the human capital production function \citep{giannola2024parental}. \textit{Second}, we use experimentally observed decisions to recover $\rho$ and classify parents into distinct behavioral types, rather than treating compensation and reinforcement as average tendencies in the data. This helps reconcile mixed empirical evidence: Studies finding reinforcement \citep[e.g.][]{aizer2012production,carneiro2023families,barrera2023compensation,berman2025sweet,giannola2024parental} and studies finding compensation \citep[e.g.][]{yi2015early} need not imply different parental motives alone, because the same parent may respond differently when the returns to concentrating resources change. 
 
The macroeconomic implications remain ambiguous. Convex returns may increase within-household inequality without necessarily increasing inequality across households if the gains to one sibling raise family resources, transfers, or future opportunities. Meanwhile, if a substantial fraction of households are highly inequality averse, their response to convexities --- compensating the lower-endowed child even more, thereby sacrificing substantial earnings --- would increase inequality between families. If inequality aversion is passed from parents to children, then this may also reduce social mobility, as inequality averse families would get increasingly stuck in low-education, low-earnings, low investment cycles.  

\newpage

\bibliography{refs.bib}

\newpage

\appendix
\section*{Online Appendix}
  
\null
\pagenumbering{arabic}
\renewcommand{\baselinestretch}{1}
\renewcommand{\thefigure} {\Alph{section}.\arabic{figure}}
\renewcommand{\thetable} {\Alph{section}.\arabic{table}}

\section{Analytical Solution of the  Allocation
Rule with Cobb-Douglas Technology}\label{app:analytical-solution}
\numberwithin{equation}{section}
\setcounter{equation}{0}

We derive the equilibrium log investment ratio under Cobb-Douglas technology ($\sigma \to 0$). The parent solves:

\begin{equation*}
\max_{x_1} \; V = \bigl[\mu \bigl(E_1(a_1,x_1)^\phi\bigr)^\rho + (1-\mu)\bigl(E_2(a_2,x_2)^\phi\bigr)^\rho\bigr]^{1/\rho}
\quad \text{s.t.} \quad x_1 + x_2 = 1,
\end{equation*} 

where $E_i = \alpha H_i^\theta$ and, under CD technology, $H_i = a_i^\gamma x_i^{1-\gamma}$. Applying the chain rule, the marginal return to investment in child $i$ is $\partial E_i/\partial x_i = \alpha\theta H_i^{\theta-1}\cdot \partial H_i/\partial x_i$, which under CD simplifies to: 

\begin{equation*}
\frac{\partial E_i}{\partial x_i} = \theta(1-\gamma)\,\frac{E_i}{x_i}.
\end{equation*} 

Substituting $x_2 = 1 - x_1$, the first-order condition $\partial V/\partial x_1 = 0$ reduces to: 

\begin{equation}
\frac{\mu\, E_1^{\phi\rho}}{x_1} = \frac{(1-\mu)\, E_2^{\phi\rho}}{x_2}.
\label{eq:foc-simplified}
\end{equation} 

Rearranging \cref{eq:foc-simplified} and substituting
$E_i = \alpha\, a_i^{\gamma\theta}\, x_i^{(1-\gamma)\theta}$ under CD
technology, letting $r \equiv x_1/x_2$ and collecting powers of $r$
yields: 

\begin{equation*}
r^{1+(\gamma-1)\theta\phi\rho} = \frac{\mu}{1-\mu}\left(\frac{a_1}{a_2}\right)^{\gamma\theta\phi\rho}.
\end{equation*} 

Taking logarithms and dividing by $\bigl[1+(\gamma-1)\theta\phi\rho\bigr]$ gives the general closed-form solution: 

\begin{equation}
\ln\!\left(\frac{x_1}{x_2}\right) = \frac{\ln\!\left(\dfrac{\mu}{1-\mu}\right) + \gamma\theta\phi\rho\,\ln\!\left(\dfrac{a_1}{a_2}\right)}{1+(\gamma-1)\theta\phi\rho}.
\label{eq:cd-allocation-general}
\end{equation}

Under the experimental design (\(\mu=0.5\)), the preference term vanishes since $\ln(\mu/(1-\mu))=\ln(1)=0$: 

\begin{equation}
\ln\!\left(\frac{x_1}{x_2}\right) = \frac{\gamma\theta\phi\rho}{1-(1-\gamma)\theta\phi\rho}\ln\!\left(\frac{a_1}{a_2}\right).
\label{eq:cd-allocation-app}
\end{equation}

Under a Cobb-Douglas production function, we can make use of the fact that:

\begin{equation}
    \ln \left(\frac{H_1}{H_2}\right) = \gamma \ln\left(\frac{a_1}{a_2}\right) + (1-\gamma) \ln \left(\frac{x_1}{x_2}\right)
\end{equation}

to rewrite \cref{eq:cd-allocation-app} over the log ratio of human capital outcomes, rather than  investments. This gives: 

\begin{equation}
\ln \left(\frac{H_1}{H_2}\right) = \frac{\gamma}{1-(1-\gamma)\theta\phi\rho}\ln\!\left(\frac{a_1}{a_2}\right).
\end{equation}

\clearpage

\section{Additional Figures and Tables}
\setcounter{table}{0}
\setcounter{figure}{0}
\subsection*{Figures}

\begin{figure}
    \centering
    \includegraphics[width=1\linewidth]{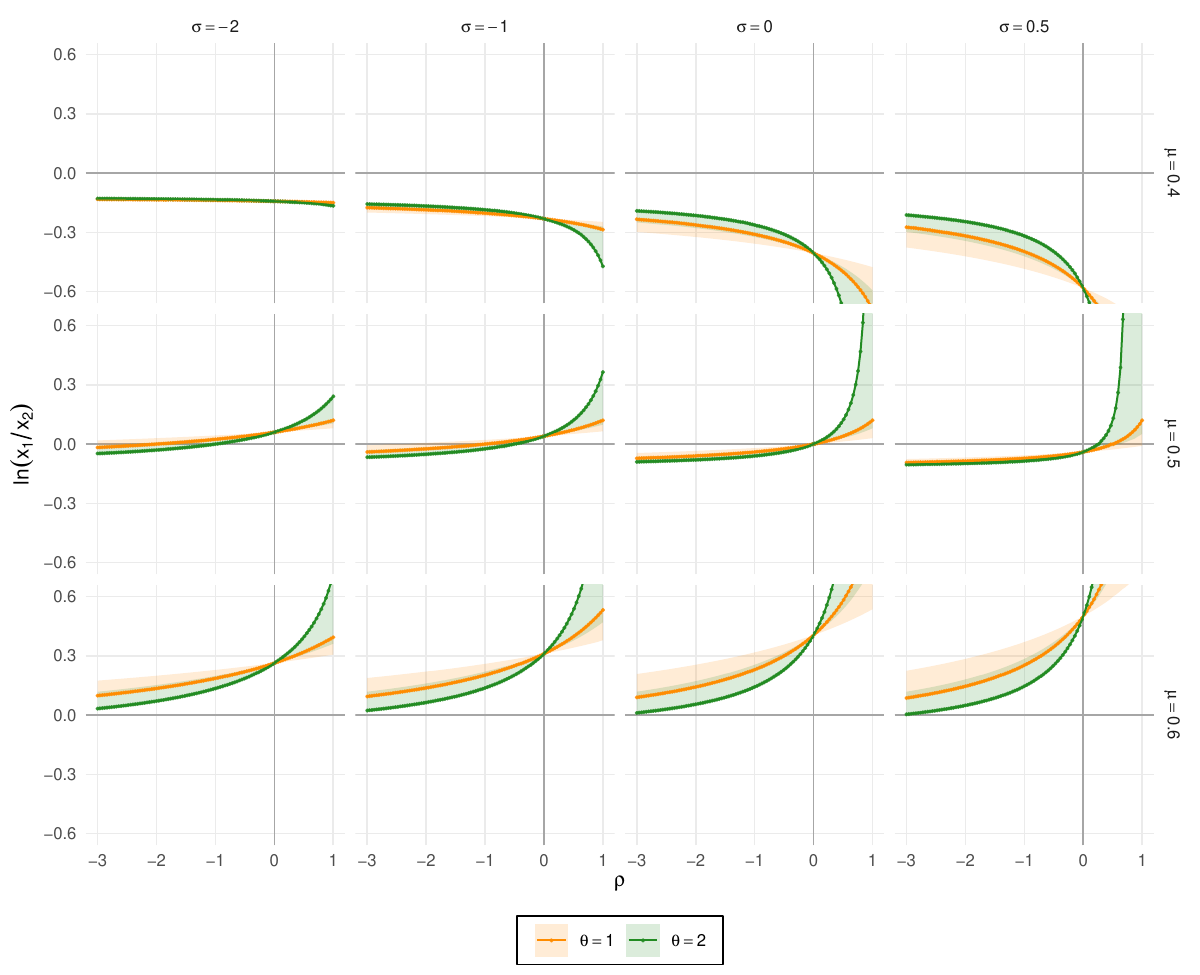} %
    \caption{Predicted allocation responses under alternative preference parameters}
    \floatfoot{\textit{Notes}:  Each panel plots the log allocation ratio $\ln(x_1/x_2)$ against parental inequality aversion $\rho$, where child~1 is the higher-ability child, so positive values indicate reinforcement and negative values compensation. 
    Columns vary the curvature $\sigma$ and rows vary the weight on the higher-ability child, $\mu$. Solid lines represent $\ln(x_1/x_2)$ when $\phi=1$, while shaded areas show how the allocation ratio varies across a range of $\phi$ values $[0.4,\,1.0]$. }
    \label{fig:relax-phi-mu}
\end{figure}

\begin{figure}
    \centering
    \includegraphics[width=0.75\linewidth]{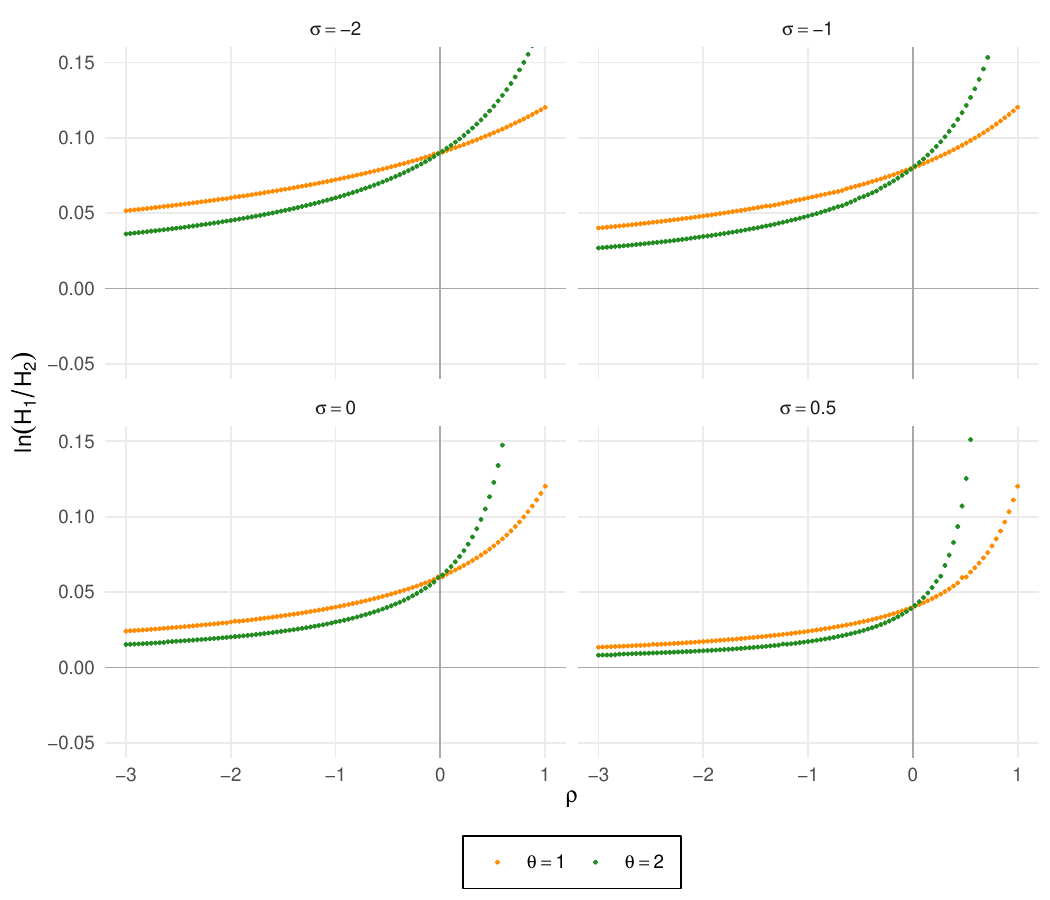}
    \caption{Predicted human capital outcomes} 
    \floatfoot{\textit{Notes:} The figure shows how $\ln{(h_{\text{1}}/h_{\text{2}}})$ varies with $\rho$ across values of $\sigma$, comparing $\theta = 1$ and $\theta = 2$.}
    \label{fig:predictions_lnHdiff}
\end{figure}

\clearpage
\subsection*{Tables}

\begin{table}[!htbp]
\centering
\caption{\label{tab:summary-estimation-sample}Summary Statistics for the Estimation Sample}
\begin{threeparttable}
\footnotesize
\setlength{\tabcolsep}{4pt}
\renewcommand{\arraystretch}{0.88}

\begin{adjustbox}{max width=\textwidth}

\begin{tabularx}{\textwidth}{Xrrrrr}
\toprule
Variable & N & Mean & SD & Min & Max\\
\midrule
\addlinespace[0.2em]
\multicolumn{6}{l}{\textbf{Parent and household}}\\
\hspace{1em}Respondent age & 243 & 42.96 & 5.56 & 33.00 & 65.00\\
\hspace{1em}Mother age & 243 & 42.56 & 5.27 & 33.00 & 57.00\\
\hspace{1em}Father age & 233 & 44.79 & 6.49 & 33.00 & 85.00\\
\hspace{1em}Household size & 244 & 4.80 & 1.37 & 2.00 & 11.00\\
\hspace{1em}Monthly household income (MXN) & 225 & 20,905 & 17,141 & 800 & 180,000\\
\hspace{1em}Respondent is mother (\%) & 246 & 68.3 &  &  & \\
\hspace{1em}Mother upper-secondary or more (\%) & 246 & 67.5 &  &  & \\
\hspace{1em}Father upper-secondary or more (\%) & 246 & 66.3 &  &  & \\
\hspace{1em}Mother tertiary or more (\%) & 246 & 26.0 &  &  & \\
\hspace{1em}Father tertiary or more (\%) & 246 & 30.5 &  &  & \\
\hspace{1em}Mother has paid occupation (\%) & 246 & 62.6 &  &  & \\
\hspace{1em}Father has paid occupation (\%) & 246 & 96.3 &  &  & \\
\hspace{1em}Two-parent household (\%) & 246 & 76.0 &  &  & \\
\hspace{1em}Mother manages household finances (\%) & 246 & 28.5 &  &  & \\
\addlinespace[0.2em]
\multicolumn{6}{l}{\textbf{Parental aspirations and beliefs}}\\
\hspace{1em}Subjective math grade, household mean & 246 & 84.95 & 9.04 & 40.00 & 100.00\\
\hspace{1em}Sibling subjective math grade gap & 246 & 5.99 & 7.76 & 0.00 & 40.00\\
\hspace{1em}Self-reported equity preference (0--10) & 246 & 7.15 & 2.77 & 0.00 & 10.00\\
\hspace{1em}Children for whom parent aspires university+ (\%) & 492 & 95.5 &  &  & \\
\hspace{1em}Children for whom parent expects university+ (\%) & 492 & 91.5 &  &  & \\
\hspace{1em}Children rated above average (\%) & 492 & 52.0 &  &  & \\
\hspace{1em}Children rated below average (\%) & 492 & 7.9 &  &  & \\
\hspace{1em}Believes university pays more than high school (\%) & 246 & 99.6 &  &  & \\
\addlinespace[0.2em]
\multicolumn{6}{l}{\textbf{Child characteristics and beliefs}}\\
\hspace{1em}Child age & 416 & 16.33 & 1.20 & 14.00 & 20.00\\
\hspace{1em}Sibling age gap & 196 & 1.46 & 0.91 & 0.00 & 4.00\\
\hspace{1em}Raven score (0--12) & 417 & 8.28 & 2.30 & 0.00 & 12.00\\
\hspace{1em}Sibling Raven score gap & 197 & 1.86 & 1.48 & 0.00 & 7.00\\
\hspace{1em}SIDE score: mother (0--1) & 408 & 0.48 & 0.07 & 0.19 & 0.81\\
\hspace{1em}SIDE score: father (0--1) & 376 & 0.49 & 0.07 & 0.17 & 0.83\\
\hspace{1em}Mixed-sex sibling pair (\%) & 197 & 43.6 &  &  & \\
\bottomrule
\end{tabularx}
\end{adjustbox}

\begin{tablenotes}[para,flushleft]
\footnotesize
\item \textit{Notes:} The sample is the final estimation sample ($N=246$ parents). Of these parents, 220 could be matched to at least one child-survey record, yielding 417 child observations; 197 households have records for both experimental children. Child-survey records are unavailable or unmatched for some children, so child-level variables are observed only for a subsample; this does not affect the parent experimental-choice sample. Child age, Raven score, and SIDE rows are calculated at the child level among matched children in the final estimation households. Sibling-gap and mixed-sex sibling-pair rows are calculated at the household level and require two matched children and nonmissing values for the relevant variable. Percentage (\%) rows report the share among the observations shown in the $N$ column. Paid occupation indicators are coded from open-text occupation responses and should be interpreted as descriptive. SIDE rows report the child self-oriented score of the Sibling Inventory of Differential Experience, separately for mothers and fathers. Scores range from 0 to 1, where values below 0.5 indicate that children report relatively more favorable treatment toward their sibling, 0.5 denotes equal treatment, and values above 0.5 indicate relatively more favorable treatment toward themselves.
\end{tablenotes}
\end{threeparttable}
\end{table}

\vspace*{\fill}
\begin{table}[htbp]
\centering
\caption{Sensitivity of $\hat{\rho}$ and Type Classifications to Alternative $\phi$ and $\mu$ Specifications}
\label{tab:rho-sensitivity-phi-mu}
\begin{adjustbox}{max width=\linewidth}

\begin{tabular}{lrrrrr}
\toprule
Specification & Rawlsian & Interior & Utilitarian & Same type as preferred & Correlation w/ preferred $\hat{\rho}$\\
\midrule
Preferred: winsorized WTP $\hat{\phi}$, $\mu=0.5$ & 31 & 189 & 26 & -- & --\\
Unwinsorized WTP $\hat{\phi}$, $\mu=0.5$ & 31 & 189 & 26 & 246 (100.0\%) & 0.998\\
$\phi=1$, $\mu=0.5$ & 32 & 193 & 21 & 226 (91.9\%) & 0.951\\
Winsorized WTP $\hat{\phi}$, SIDE-based $\hat{\mu}$ & 36 & 187 & 23 & 234 (95.1\%) & 0.974\\
\bottomrule
\end{tabular}
\end{adjustbox}
\par\smallskip
\begin{minipage}{\linewidth}
\footnotesize
\textit{Notes:} $N = 246$ for all specifications. The preferred specification --- chosen for its goodness-of-fit metrics (cf. \Cref{app:model-fit}) --- uses the winsorized WTP estimate of $\phi$ and fixes $\mu=0.5$. Each alternative changes only the indicated specification of $\phi$ or $\mu$. `Same type as preferred' reports the number and percentage of parents whose Rawlsian, Interior, or Utilitarian classification matches the preferred specification. The final column reports Spearman's rank correlation between the alternative and preferred $\hat{\rho}$ estimates.
\end{minipage}
\end{table}
\vspace*{\fill}

\begin{landscape}
\begin{table}[!htbp]
\centering
\small
\sisetup{group-separator={,}, group-minimum-digits=4}
\begin{threeparttable}
\caption{Hypothesis Tests in the Final Estimation Sample}
\label{tab:hypothesis-tests-final}

\resizebox{\textwidth}{!}{%

\begin{tabular}{lllrrrrrrr}
\toprule
Hypothesis & Outcome & Term & Respondents & Obs. & Estimate & SE & p-val (two-sided) & p-val (one-sided) & Within $R^2$\\
\midrule
H1a & $x_{\text{high}}$ & $\sigma$ & \num{246} & \num{1476} & \num{-0.236} & \num{0.049} & $<0.001$ & $<0.001$ & \num{0.024}\\
H1b & reinforce & $\sigma$ & \num{246} & \num{1476} & \num{-0.115} & \num{0.008} & $<0.001$ & $<0.001$ & \num{0.104}\\
H2a & $x_{\text{high}}$ & $\theta$ & \num{246} & \num{1476} & \num{1.882} & \num{0.128} & $<0.001$ &  & \num{0.252}\\
H2b & reinforce & $\theta$ & \num{246} & \num{1476} & \num{0.260} & \num{0.023} & $<0.001$ &  & \num{0.114}\\
H3 & payoffmax & $\theta$ & \num{246} & \num{1476} & \num{-0.297} & \num{0.024} & $<0.001$ & $<0.001$ & \num{0.173}\\
\bottomrule
\end{tabular}
}

\begin{tablenotes}[flushleft]
\footnotesize
\item \parbox{\textwidth}{%
\textit{Notes}: All five tests use the final estimation sample of \num{246} respondents and \num{1476} scenario observations. Model-derived $\hat{\rho}$-type restrictions are reported in \Cref{tab:reduced-form-robust}. One-sided $p$-values are reported for the preregistered directional H1 and H3 tests; H2 is non-directional. All models include display-order fixed effects, and standard errors are clustered at the level of the parent.%
}
\end{tablenotes}

\end{threeparttable}
\end{table}
\end{landscape}

\newpage
\begingroup
\small
\sisetup{group-separator={,}, group-minimum-digits=4}
\begin{longtblr}[
  caption = {Reduced-Form Robustness Checks},
  label   = {tab:reduced-form-robust},
  note{}  = {\textit{Notes:} Column (1) includes all \num{246} respondents in the final estimation sample, for reference. In column (2), Panels A and B exclude Rawlsian parents, while Panel C excludes Utilitarian parents ($\hat{\rho}\geq 1-10^{-6}$). Standard errors are clustered at the parent level. Stars reflect two-sided p-values. \\ * $p<0.1$, ** $p<0.05$, *** $p<0.01$.}
]{
  colspec = {Q[l,4.2cm] Q[c,3.1cm] Q[c,3.1cm]},
  rowsep = 0.5pt,
  colsep = 3pt,
  rowhead = 1,
}
\toprule
 & {(1) All respondents} & {(2) Model-derived restriction} \\
\midrule
\SetCell[c=3]{l}{\textbf{Panel A. H1a:} $x_{\text{high}}$} \\
\midrule
$\sigma$ & \num{-0.236}*** & \num{-0.402}*** \\
 & (\num{0.049}) & (\num{0.046}) \\
Obs. & \num{1476} & \num{1290} \\
Households & \num{246} & \num{215} \\
Dep. var. mean & \num{5.257} & \num{5.650} \\
Within $R^2$ & \num{0.024} & \num{0.054} \\
Display-order FE & \checkmark & \checkmark \\
\midrule
\SetCell[c=3]{l}{\textbf{Panel B. H1b:} reinforce} \\
\midrule
$\sigma$ & \num{-0.115}*** & \num{-0.131}*** \\
 & (\num{0.008}) & (\num{0.009}) \\
Obs. & \num{1476} & \num{1290} \\
Households & \num{246} & \num{215} \\
Dep. var. mean & \num{0.483} & \num{0.553} \\
Within $R^2$ & \num{0.104} & \num{0.119} \\
Display-order FE & \checkmark & \checkmark \\
\midrule
\SetCell[c=3]{l}{\textbf{Panel C. H3:} payoffmax} \\
\midrule
$\theta$ & \num{-0.297}*** & \num{-0.351}*** \\
 & (\num{0.024}) & (\num{0.023}) \\
Obs. & \num{1476} & \num{1320} \\
Households & \num{246} & \num{220} \\
Dep. var. mean & \num{0.611} & \num{0.578} \\
Within $R^2$ & \num{0.173} & \num{0.231} \\
Display-order FE & \checkmark & \checkmark \\
\bottomrule
\end{longtblr}
\endgroup

\begin{table}[!htbp]
\centering
\small
\sisetup{group-separator={,}, group-minimum-digits=4}
\begin{talltblr}[
  caption = {Effect of Convexity by $\rho$ Type},
  label   = {tab:h2-rho-type},
  note{} = {\textit{Notes:} This table reports rho type-specific marginal effects of $\theta$, estimated from interactions between $\theta$ and $\hat{\rho}$ type. Dependent-variable means are unconditional means over the six scenario observations within each $\hat{\rho}$ type. %
  Type counts are Interior = \num{189}, Rawlsian = \num{31}, and Utilitarian = \num{26} respondents. Rho types are defined in \Cref{subsec-rho}. All specifications include display-order fixed effects, with standard errors clustered at the parent level.\\ * $p<0.1$, ** $p<0.05$, *** $p<0.01$.}
]{
  colspec = {Q[l,4.4cm] Q[c,2.25cm] Q[c,2.25cm]},
  rowsep = 1pt,
  colsep = 2pt,
}
\toprule
Outcome: & $x_{\text{high}}$ & reinforce \\
 & (1) & (2) \\
\midrule
\SetCell[c=3]{l} Marginal effect of $\theta$ by type: & & \\
\addlinespace
Rawlsian & \num{-0.021} & \num{0.000} \\
 & (\num{0.068}) & (\num{0.003}) \\
Interior & \num{2.019}*** & \num{0.288}*** \\
 & (\num{0.150}) & (\num{0.028}) \\
Utilitarian & \num{3.157}*** & \num{0.372}*** \\
 & (\num{0.208}) & (\num{0.046}) \\
\addlinespace
\SetCell[c=3]{l} Mean dependent variable by $\hat{\rho}$ type: & & \\
\addlinespace
Rawlsian & \num{2.527} & \num{0.000} \\
Interior & \num{5.411} & \num{0.517} \\
Utilitarian & \num{7.391} & \num{0.814} \\
\midrule
Obs. & \num{1476} & \num{1476} \\
Respondents & \num{246} & \num{246} \\
Within $R^2$ & \num{0.297} & \num{0.131} \\
Display-order FE & \checkmark & \checkmark \\
\bottomrule
\end{talltblr}
\end{table}

\begin{table}[!htbp]
\centering
\small
\sisetup{group-separator={,}, group-minimum-digits=4}
\begin{talltblr}[
  caption = {Reduced-Form Robustness Checks: All Experimental Parents},
  label   = {tab:reduced-form-all-parents},
  note{}  = {\textit{Notes:} This table replicates the main reduced-form specifications presented in \Cref{tab:reduced-form}, including parents who did not pass the practice-round comprehension check and parents whose allocations never vary. No $\hat{\rho}$-based sample restriction is imposed. The dependent variable is reported in column header. Standard errors clustered at the parent level are shown in parentheses. Stars are based on two-sided p-values. \\
  * $p<0.1$, ** $p<0.05$, *** $p<0.01$.}
]{
  colspec = {Q[l,3.2cm] Q[c,1.85cm] Q[c,1.85cm] Q[c,1.85cm] Q[c,1.85cm] Q[c,1.85cm]},
  rowsep = 1pt,
  colsep = 2pt,
  hline{7} = {1-6}{solid, 0.05em, black},
}
\toprule
Outcome: & $x_{\text{high}}$ & reinforce & $x_{\text{high}}$ & reinforce & payoffmax \\
 & (1) & (2) & (3) & (4) & (5) \\ 
\midrule
$\sigma$ & \num{-0.185}*** & \num{-0.106}*** &  &  &  \\
 & (\num{0.046}) & (\num{0.008}) &  &  &  \\
$\theta$ &  &  & \num{1.676}*** & \num{0.243}*** & \num{-0.294}*** \\
 &  &  & (\num{0.120}) & (\num{0.021}) & (\num{0.025}) \\
Obs. & \num{1668} & \num{1668} & \num{1668} & \num{1668} & \num{1668} \\
Households & \num{278} & \num{278} & \num{278} & \num{278} & \num{278} \\
Dep. var. mean & \num{5.199} & \num{0.471} & \num{5.199} & \num{0.471} & \num{0.581} \\
Within $R^2$ & \num{0.017} & \num{0.093} & \num{0.214} & \num{0.104} & \num{0.156} \\
Display-order FE & \checkmark & \checkmark & \checkmark & \checkmark & \checkmark \\
\bottomrule
\end{talltblr}
\end{table}

\newpage

\clearpage
\section{Practice Rounds and Comprehension
Check}\label{app:comprehension}

The two practice rounds used a store-allocation framing designed to be immediately intuitive while holding all preference parameters constant by construction. Parents were told the following (paraphrased from the standardized enumerator script):

\begin{quote}
\emph{Imagine you have nine tokens to spend between two supermarkets:
Soriana (\textcolor{orange}{orange}) and Walmart (\textcolor{Green}{green}). Both stores are located side by
side and sell exactly the same products at exactly the same prices. The only difference is that one store values the tokens more: Spending five tokens at that store will buy more items than spending five tokens at the other. Your task is to decide how to distribute your tokens between the two stores in the way that benefits you most. Please move the slider accordingly.}
\end{quote}

Because the two stores are identical in every dimension except token value, i.e. same goods, same prices, no brand preference, the framing ensures ($\mu = 0.5$) (no reason to prefer one store) and ($\phi = 1$) (items are homogeneous, so each additional item carries equal marginal value) by design. Both practice rounds used parameters ($\sigma = 1$) and ($\theta = 2$), a setting in which the payoff-maximizing allocation, all nine tokens to the higher-value store (\textcolor{orange}{Soriana}), is uniquely determined and requires no calculation beyond understanding that a higher token multiplier yields more items. A parent who placed all nine tokens on Soriana showed that they had correctly understood the mapping from token allocations to payoffs; we use this as a binary comprehension indicator.

\clearpage
\section{Power Analysis}\label{app:power}
\setcounter{table}{0}
\setcounter{figure}{0}
We pre-registered an analytical power analysis to determine the minimum sample size needed to detect our hypothesized effects at various effect sizes while minimizing Type II error.

For each parent we simulate optimal token allocations across six scenarios (two levels of $\theta \in \{1, 2\}$ and three levels of $\sigma \in \{-2, 0, 0.5\}$) using our theoretical model and adding Gumbel-distributed choice noise with scale parameter
$\sigma_{\text{noise}} = 0.5$ tokens to reflect real-world deviations from theoretically optimal behavior due to rounding (integer token constraint), attention lapses, and cognitive limitations. We construct three outcome measures: (i) Continuous allocation to the high-endowment child $(x_{\text{high}})$, (ii) reinforcement indicator (allocating $\geq 5$ tokens to high-endowment child), and (iii) payoff-maximization indicator. We test the directional hypotheses in \Cref{tab:power-analysis} using one-sided tests at significance level $\alpha = 0.05$, with standard errors clustered at the individual level to account for within-individual correlation. The non-directional hypotheses are tested with two-sided tests. The expected effect sizes (d) vary between 0.2 and 0.3.

\begin{table}[h!t]
    \centering
    \footnotesize
    \begin{tabular}{lccc}
        \toprule
        \textbf{Hypothesis} & \textbf{Power (\%)} & \textbf{Power (\%)} & \textbf{Test} \\
        \textbf{} & \textbf{$n=50$} & \textbf{$n=240$} & \\
        \midrule
        H1a: Substitutability $\to$ allocations (d=0.2) & 99.9 & 100.0 & $x_{\text{high}} \sim \sigma$ (continuous) \\
        H1b: Substitutability $\to$ reinforcement (d=0.3) & 100.0 & 100.0 & reinforce $\sim \sigma$ (binary) \\
        H2: Convexity $\to$ reinforcement (d=0.2) & 27.1 & 92.6 & reinforce $\sim \theta$ (binary) \\
        H3: Payoff maximization (d=0.3)& 74.3 & 100.0 & payoffmax $\sim \theta$ (binary) \\
        \bottomrule
    \end{tabular}
    \caption{Statistical power for hypothesis tests based on observed effects from structural model simulations}
    \label{tab:power-analysis}
\end{table}
\normalsize
\newpage

\section{Model Fit Details}\label{app:model-fit}
\setcounter{table}{0}
\setcounter{figure}{0}

\paragraph*{Estimated Stage-Two GMM weights.} Relative to the average scenario, the stage-two weighting matrix assigns less direct weight to the linear-return scenarios and more direct weight to the convex-return scenarios (see \Cref{tab:stage2-weights}). The largest normalized diagonal weight is assigned to the  $\theta=2,\sigma=0.5$ scenario, followed by  $\theta=2,\sigma=-2$ scenario, and the  $\theta = 2, \sigma = 0$ scenario. This indicates that, after accounting for the cross-parent residual covariance structure, the convex-return moments receive greater influence in the weighted minimum-distance criterion.

\begin{table}[htbp]
\centering
\caption{Diagonal Weights from the Stage-Two Population GMM Weighting Matrix}
\label{tab:stage2-weights}
\begin{threeparttable}

\begin{tabular}{crr}
\toprule
$\theta$ & $\sigma$ & Normalized diagonal weight\\
\midrule
1 & $-2.0$ & 0.941\\
1 & $0.0$ & 0.908\\
1 & $0.5$ & 0.448\\
2 & $-2.0$ & 1.244\\
2 & $0.0$ & 1.046\\
2 & $0.5$ & 1.412\\
\bottomrule
\end{tabular}
\begin{tablenotes}[flushleft]
\footnotesize
\item \textit{Notes:} Weights are normalized by the mean diagonal entry of the stage-two population GMM weighting matrix. Values above one indicate scenarios receiving greater direct weight than the average scenario.
\end{tablenotes}
\end{threeparttable}
\end{table}

\paragraph*{Individual-level fit ($n = 246$ parents).} The root mean squared moment error
(RMSME $=\sqrt{\mathrm{SSR}_{\mathrm{unweighted}}/6}$) measures the average per-scenario prediction error in tokens for each parent.

\begin{table}[htbp]
\centering
\caption{Distribution of per-parent RMSME}
\label{tab:parent_rmsme}

\begin{tabular}{lr}
\toprule
Statistic & Value\\
\midrule
Mean RMSME & 1.11 tokens\\
Median RMSME & 1.17 tokens\\
IQR & $[0.68,\ 1.37]$\\
Range & $[0.14,\ 3.71]$\\
\bottomrule
\end{tabular}
\end{table}

RMSME is below one token for 43.1\% of parents, between one and 1.5 tokens for another 42.7\%, and at least 1.5 tokens for the remaining 14.2\%.

\paragraph*{Scenario-specific performance.} 
The scenario-specific marginal performance reveals that the model fits convex-return scenarios substantially better than linear-return scenarios. Mean absolute error is $0.59$ when $\theta=2$, compared with $1.09$ when $\theta=1$ (cf. \Cref{tbl-fit-by-theta}). Across production technologies, MAE is lowest under low substitutability ($\sigma=-2$, MAE $=0.69$), followed by the Cobb–Douglas case ($\sigma=0$, MAE $=0.82$), and highest under high substitutability ($\sigma=0.5$, MAE $=1.02$). The corresponding mean residuals are $-0.23$, $-0.07$, and $-0.25$ (cf. \Cref{tbl-fit-by-sigma}). Thus, the model over-predicts allocation to the higher-endowed child on average in all three substitutability conditions, with the smallest mean bias in the Cobb–Douglas condition.

\paragraph*{Fatigue check.} The MAE computed by the within-parent display order in which scenarios were presented (orders 1 through 6) varies non-monotonically between $0.69$ and $0.91$ tokens, with no systematic trend across the sequence. We find no evidence that fatigue or inattention contaminate the structural estimates.

\paragraph*{Hansen's J-test.} The overidentification test evaluates whether the six moment conditions are jointly consistent with the data for each parent. Under the specification, $J_i = \mathbf{m}_i(\hat{\rho})' W \mathbf{m}_i(\hat{\rho})$ follows a $\chi^2_5$ distribution. \Cref{tbl-j-test} reports the distribution across the sample.

\begin{table}[htbp]
\centering
\caption{Distribution of Hansen's $J$-statistics ($n = 246$)}
\label{tbl-j-test}

\begin{tabular}{lr}
\toprule
Statistic & Value\\
\midrule
Mean $J$ & $5.73$\\
Median $J$ & $2.55$\\
Critical value (5\%, $df=5$) & $11.07$\\
Mean $p$-value & $0.61$\\
Median $p$-value & $0.77$\\
Rejection rate at 5\% & $13.0\%$\\
\bottomrule
\end{tabular}
\end{table}

The model fits the typical parent well. The median $J$-statistic is $2.55$, with a median $p$-value of $0.77$, indicating little evidence against the overidentifying restrictions for the median parent. The sample mean $J$ of $5.73$ is also well below the $\chi^2_5$ critical value of $11.07$, and the average $p$-value of $0.61$ points in the same direction. At the conventional $5\%$ level, the test rejects for 32 of 246 parents, roughly one in eight. Note that the broader GMM literature cautions that feasible optimal weighting can have non-negligible finite-sample consequences because the weighting matrix $W=S^{-1}$ is estimated rather than known \citep{hansen1996finite, altonji1996small}. This concern is especially relevant here, where $J_i$ is constructed parent by parent from only six scenario residuals. We therefore treat the $\chi^2_5$ distribution as a reference approximation rather than an exact finite-sample benchmark. The 13\% rejection rate points to localized tension with the model's restrictions, but does not by itself imply widespread misspecification.

\noindent
\begin{minipage}[t]{0.47\textwidth}
\centering
\captionof{table}{Marginal fit by $\theta$}
\label{tbl-fit-by-theta}
\renewcommand{\arraystretch}{0.6}

\begin{tabular}{rrrr}
\toprule
$\theta$ & MAE & RMSE & Mean residual\\
\midrule
1 & 1.092 & 1.51 & $-0.599$\\
2 & 0.594 & 0.99 & $+0.236$\\
\bottomrule
\end{tabular}
\end{minipage}%
\hfill
\begin{minipage}[t]{0.47\textwidth}
\centering
\captionof{table}{Marginal fit by $\sigma$}
\label{tbl-fit-by-sigma}
\renewcommand{\arraystretch}{0.6}

\begin{tabular}{rrrr}
\toprule
$\sigma$ & MAE & RMSE & Mean residual\\
\midrule
$-2$ & 0.689 & 1.11 & $-0.231$\\
$0$ & 0.817 & 1.15 & $-0.065$\\
$0.5$ & 1.022 & 1.52 & $-0.249$\\
\bottomrule
\end{tabular}
\end{minipage}

\end{document}